\documentclass[journal=jpcbfk,manuscript=article]{achemso}
\usepackage{longtable}
\usepackage{multirow}
\usepackage{amsmath}
\usepackage[usenames,dvipsnames]{color}
\usepackage{soul}
\usepackage{graphicx}
\usepackage{amsmath,amssymb}
\usepackage{caption}
\usepackage{color}
\usepackage{dcolumn}
\usepackage{bm}
\usepackage{float}
\usepackage{subcaption}
\usepackage{textcomp}
\usepackage{url}
\usepackage{booktabs}
\usepackage{microtype}
\usepackage{tabularx}
\usepackage{siunitx}

\usepackage[normalem]{ulem}
\setcitestyle{square,numbers,sort&compress}

\makeatletter

\author{Jiri K\"aser} \affiliation[University of Basel]{Department of
  Mathematics and Computer Science, University of Basel,
  Spiegelgasse 1, CH-4051 Basel, Switzerland}

\author{Kai T\"opfer} \affiliation[University of Basel]{Department of
  Chemistry, University of Basel, Klingelbergstrasse 80 , CH-4056
  Basel, Switzerland}

\author{Markus Meuwly} \affiliation[University of Basel]{Department of
  Chemistry, University of Basel, Klingelbergstrasse 80 , CH-4056
  Basel, Switzerland} \altaffiliation{Department of Chemistry, Brown
  University, Providence, RI 02912, USA} \email{m.meuwly@unibas.ch}

\title{Diffusion and Spectroscopy of H$_2$ in Myoglobin}

\begin{document}
\date{\today}

\begin{abstract}
The diffusional dynamics and vibrational spectroscopy of molecular
hydrogen (H$_2$) in myoglobin (Mb) is characterized. Hydrogen has been
implicated in a number of physiologically relevant processes,
including cellular aging or inflammation. Here, the internal diffusion
through the protein matrix was characterized and the vibrational
spectroscopy was investigated using conventional empirical energy
functions and improved models able to describe higher-order
electrostatic moments of the ligand. H$_2$ can occupy the same
internal defects as already found for Xe or CO (Xe1 to Xe4 and
B-state). Furthermore, 4 additional sites were found, some of which
had been discovered in earlier simulation studies. The vibrational
spectra using the most refined energy function indicate that depending
on the docking site the spectroscopy of H$_2$ differs. The maxima of
the absorption spectra cover $\sim 20$ cm$^{-1}$ which are indicative
of a pronounced effect of the surrounding protein matrix on the
vibrational spectroscopy of the ligand. Electronic structure
calculations show that H$_2$ forms a stable complex with the heme-iron
(stabilized by $\sim -12$ kcal/mol) but splitting of H$_2$ is unlikely
due to a high activation energy ($\sim 50$ kcal/mol).
\end{abstract}

\section{Keywords}
Myoglobin, Vibrational Spectroscopy, MD simulations, H$_2$ ligand,
QM/MM, Advanced Electrostatics, Stark Shift

\section{Introduction}
The interaction of Myoglobin (Mb) with small molecules is of profound
interest from a physiological perspective. Myoglobin is structurally
related to one of the two subunits $(\alpha, \beta)$ of Hemoglobin
(Hb) both of which bind, store, and transport diatomic ligands (O$_2$,
NO, CO). Due to their high biological relevance, both proteins
interacting with all three diatomics have been intensely studied
experimentally
\cite{austin:1975,cornelius:1983,martin:1983,steinbach:1991,schotte:2003,barends:2015,barends:2024}
and computationally
\cite{MM.mbco:2004,MM.mbno:2002,MM.mbno:2015,MM.mbno:2016,cohen:2006}.\\

\noindent
The tertiary structure of Mb is characterized by a number of internal
cavities. These were first mapped out by pressurizing the protein with
xenon gas \cite{tilton:1984}. The physiological role these packing
defects may play indicate that blocking sites "B" and/or "Xe4" have important
consequences for the overall rates of oxygen binding to
Mb \cite{olson:2007}. The dynamics of ligands between these sites were
also investigated using computer simulations \cite{MM.mb:2023}.
Interestingly, proteins other than Mb also display such cavities,
including neuroglobin or truncated
hemoglobin \cite{milani:2004,MM.trhb:2012,brunori:2007,MM.ngb:2009}.
Hence, such packing defects may be functionally relevant and
characterizing them and the dynamics between them for various ligands
is of general and fundamental interest.\\

\noindent
Molecular hydrogen, H$_2$, has been reported to play physiologically
relevant roles in cell protection by reducing hydroxyl radicals
\cite{ohsawa:2007}, showed therapeutic effects in carcinoma after
hyperbaric hydrogen therapy in mice \cite{dole:1975}, and has been
used as a medical therapeutic gas to treat brain disorders
\cite{wu:2023}. Furthermore, through its antioxidative effect, H$_2$
maintains genomic stability, mitigates cellular aging, influences
histone modification, telomere maintenance, and proteostasis. In
addition, the diatomic may prevent inflammation and regulate the
nutrient-sensing mTOR system, autophagy, apoptosis, and mitochondria,
which are all factors related to aging \cite{fu:2022}.  Hence, H$_2$
is expected to play various roles in the human body.\\

\noindent
Myoglobin and other heme-based proteins are known to interact with
small molecules, in particular diatomics. Gas inhalation as disease
therapy has been investigated and heme-containing proteins - in
particular cytochrome c oxidase - have been found to be primary
targets for small molecules such as O$_2$, NO, CO or H$_2$S
\cite{szabo:2007,kajimura:2010}. Furthermore, H$_2$ did not reduce the
oxidized heme in cytochrome c which implicated that the primary target
for H$_2$ appeared to be different from cytochrome c oxidase
\cite{ohsawa:2007}.  Interestingly, combined therapy with H$_2$ and CO
demonstrated enhanced therapeutic effects \cite{nakao:2010}. This is
akin to hyperbaric hydrogen treatment which used 2.5\% O$_2$ combined
with 97.5\% H$_2$ and showed regression of skin tumors in mice
\cite{dole:1975}.\\

\noindent
One of the less studied interaction partners of Mb is the hydrogen
molecule, H$_2$. Due to its small size and the fact that it is
electrically neutral, it can be expected to diffuse easily into and
within the protein. On the other hand, H$_2$ has an appreciable
electric quadrupole moment \cite{orcutt:1963}.  Hence, from a
spectroscopic perspective it may behave in a similar fashion as CO
which is also electrically neutral, with a rather small permanent
dipole moment but a sizable molecular quadrupole
\cite{maroulis:2001}.\\

\noindent
The present work characterizes the ligand diffusion and vibrational
spectroscopy of H$_2$ in Mb from atomistic simulations. Furthermore,
the binding and chemistry of H$_2$ attached to the heme-iron is
considered. First, the computational methods are presented. This is
followed by characterizing the interaction between H$_2$ and the
heme-group, the structural dynamics of H$_2$ in the protein and the
vibrational spectroscopy in particular internal pockets. Finally,
conclusions are drawn.\\

\section{Material and Methods}
\subsection{Molecular Dynamics Simulations and Analysis}
Molecular Dynamics (MD) simulations of myoglobin and one or five H$_2$
molecules were performed using the CHARMM
\cite{Charmm-Brooks-2009,MM.charmm:2024} molecular simulation package
starting from the X-ray structure 1MBC \cite{kuriyan:1986} prepared as
described in previous work \cite{MM.mb:2009}. H$_2$ was initially
inserted within the known pockets for xenon atoms in myoglobin
\cite{tilton:1984,cohen:2006,MM.mb:2023}. The simulations were carried
out in a cubic box of size $(61.3)^3$ \AA\/$^3$ with explicit TIP3P
water \cite{jorgensen-jcp-83} and including 29 K$^+$ and Cl$^-$
ions. Bond lengths involving H-atoms were constrained using SHAKE
\cite{SHAKE-Gunsteren-1997} except for the H$_2$ molecule itself. A
cutoff of 12 \AA\/ with switching at 10 \AA\/ was used for non-bonded
interactions \cite{Steinbach1994}. The system was initially heated in
the $NVT$ simulation to $303.15$ K for 40 ps, followed by 40 ps
equilibration simulation in the $NpT$ ensemble at $p = 1$ bar using
the leap-frog integrator and a Hoover thermostat
\cite{Hoover1985}. Subsequently, production runs were performed for 5
ns again using the $NpT$ ensemble.\\

\noindent
To identify internal localization sites for the H$_2$ ligand,
candidate sequences of simulation trajectories were selected. The
H$_2$ ligand was considered occupying a pocket if the H$_2$ center of
mass remained within 5 \AA\/ of the pocket center for a time
$\tau_{\rm dwell}$. The dwell time chosen was $\tau_{\rm dwell} = 20$
ps but is largely arbitrary.  Iterating over all candidate sequences
the distances between the H$_2$ center of mass and all C$_\alpha$
atoms were computed for the trajectories. Within the search radius of
10 \AA\/ a set of C$_\alpha$ atoms in amino acid residues was selected
so that that their geometric center overlaps best with the average
center of mass of H$_2$ moving inside the pocket.  The procedure was
repeated with additional simulations and candidate sequences, if no
clear pocket could be identified.  Pockets Xe1 to Xe4, the B-state,
and pockets 6 to 9 identified by following this procedure are shown in
Figure \ref{fig_pocket_positions} and their definition by the residue
number and name are documented in Table S1. \\

\noindent
The line shape $I(\omega)$ of the power spectra for H$_2$ in myoglobin
are obtained via the Fourier transform of the distance-distance
autocorrelation function from the H$_2$ separation $d(t)$
\begin{equation}
  I(\omega) n(\omega) \propto Q(\omega) \cdot \mathrm{Im}\int_0^\infty
  dt\, e^{i\omega t} 
  \left \langle 
  r(t)
  \cdot r(0) \right \rangle
\label{eq:powerspectra}
\end{equation}
where $r$ is the H$_2$ bond length. A quantum correction factor
$Q(\omega) = \tanh{(\beta \hbar \omega / 2)}$ was applied to the
results of the Fourier transform \cite{marx:2004}. This procedure
yields H$_2$ vibrational signals at the correct frequencies but not
the absolute intensities as observed in experimentally measured
vibrational spectra.  \\

\subsection{The Energy Function}
The MD simulations were carried out using the all-atom force field
CHARMM36 (CGenFF) \cite{CHARMMFF-ALL36-Guvench2011}, the corresponding
TIP3P water model \cite{TIP3P-Jorgensen-1983} and the Lennard-Jones
(LJ) parameter to describe the non-bonded van-der-Waals interaction
for the K$^+$ and Cl$^-$ ions \cite{mackerell:2015}. For H$_2$ the
bonded interaction was either a harmonic potential with a force
constant of $k = 350$ kcal/mol/\AA$^2$ and $r_e = 0.7414$ \AA\/ or a
Morse potential fitted to energies determined at the
CCSD(T)/aug-cc-pVQZ level of theory using the Gaussian16 program code
\cite{g09}.  For this, the H$_2$ bond was scanned between 0.5 and
1.5\,\AA\/ in steps of 0.01\,\AA\/ and the energies were represented
as $V(r) = D_e\cdot (1 - \mathrm{e}^{-\beta \cdot (r - r_e)})^2$. The
fit yielded $D_e = 111.76$\,kcal/mol, $r_e = 0.7477$\,\AA\/ and $\beta
= 1.9487$ \AA\/$^{-1}$. The Lennard-Jones parameters for H$_2$ were
those from the literature \cite{h2lj:2021} which were fitted for
accurate H$_2$ gas and the interface biomolecules. For the H$_2$
electrostatics a minimal distributed charge model was developed as
described below.\\

\noindent
To validate the H$_2$ stretch potential the anharmonic frequency was
determined from solving the nuclear Schr\"odinger equation based on a
discrete variable representation. The fundamental transition was found
at $\nu = 4202.4$\,cm$^{-1}$ compared with the experimentally reported
value of 4161.1\,cm$^{-1}$ for the rotationless transition
\cite{stoicheff:1957}. For the harmonic potential the frequency is at
$\omega = 4047.0$\,cm$^{-1}$.\\

\subsection{Electronic Structure Calculations}
For the potential energy surface (PES) for H$_2$ interacting with the
heme unit, electronic structure calculations using the ORCA program
were carried out \cite{orca:2020}. The PES was scanned along the H$_2$
bond $r$ and the $z-$direction between the Fe atom and H$_2$ center of
mass. A mixed quantum mechanics/molecular mechanics (QM/MM) approach
was adopted to include electrostatic interactions between the
surrounding protein and the His-Heme-H$_2$ subsystem. For this, all
protein atoms were assigned their CGenFF charges and the
His-Heme-H$_2$ subsystem was treated at the rPBE/def2-TZVP level of
theory including D4 dispersion corrections
\cite{ahlrichs:2005,weigend:2006,grimme:d4}. First, the structure of
H$_2$ was optimized with otherwise constrained atom positions. Then,
scans along directions $z$ and $r$ were carried out, see Figure S1.
The $z$-direction was set as the normalized vector
from the Fe to the center of mass of H$_2$ whereas $r$-direction was
defined between both H atoms, is corrected to be orthogonal to the
$z$-vector also considering the optimized position and rotation of
H$_2$ on Fe in Myoglobin. The center of mass of H$_2$ was shifted
along the $z$-direction to match distances to the Fe atom from 0.28
\AA\/ to 3.18 \AA\/ in 0.1 \AA\/ steps. From the shifted center of
mass the H$_2$ atoms were set apart from 0 to 3.2 \AA\/ along the
$r$-direction in 0.2 \AA\/ steps, conserving the center of mass.

\subsection{MDCM Model for H$_2$}
With a standard force field-based energy function, H$_2$ only
interacts through van-der-Waals interactions with its environment due
to its neutrality and vanishing molecular dipole moment. This,
however, does not account for its nonvanishing quadrupole moment. To
capture this, the electrostatic potential (ESP) of the neutrally
charged H$_2$ molecule is reproduced by the minimal distributed charge
model (MDCM) \cite{MM.mdcm:2017} using 3 off-centered charges per
hydrogen atom.  The off-centered charges are restricted along the bond
axis and symmetric to the horizontal mirror plane perpendicular to the
H$_2$ bond. The reference ESP is computed at the CCSD(T)/aug-cc-pVQZ
level of theory using Gaussian16 at the H$_2$ equilibrium conformation
at the CCSD/Aug-cc-pVQZ level of theory ($r_e=0.7424$\,\AA) and
represented in a cube file format with Gaussian16's default coarse
grid resolution \cite{g09}.\\

\begin{figure}
\includegraphics[width=0.90\linewidth]{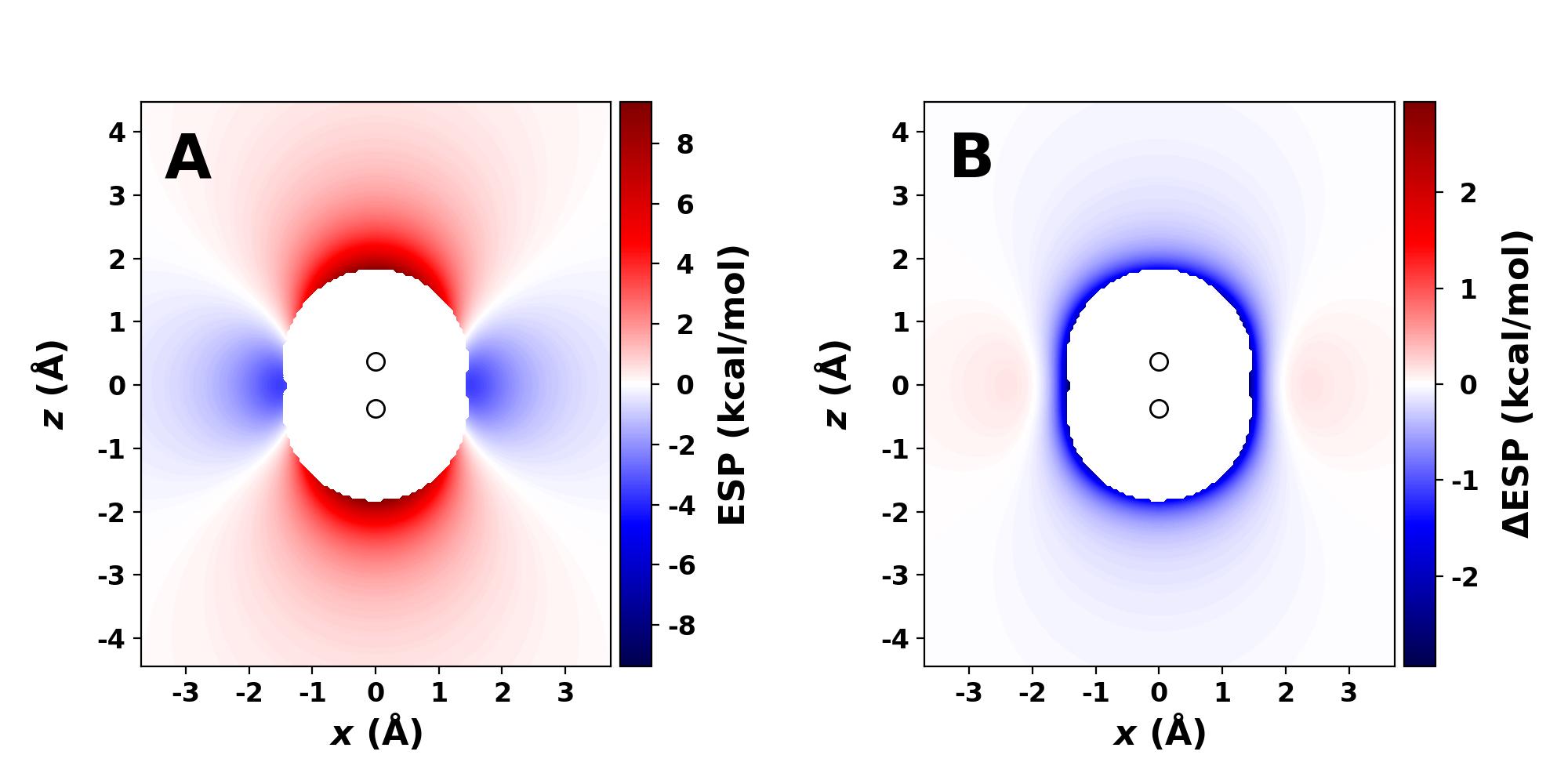}
\caption{{\bf H$_2$ ESP fit.}  Panel A: Reference ESP contour plot of
  H$_2$ along the $(x,z)$-plane going through the molecule.  H$_2$ is
  at equilibrium bond length and the ESP is only shown for for grid
  points with distances larger than $1.44$\,\AA\/ of the closest
  hydrogen atom.  Panel B: ESP difference contour plot between
  reference and model ESP.  The open black circles mark the positions
  of the hydrogen atoms.}
\label{fig_esp_h2}
\end{figure}

\noindent
The off-centered charge displacements and amplitudes are optimized to
best fit the ESP grid points in the range of 1.44 ($1.2 \cdot
r_\mathrm{vdW}$) to 2.64\,\AA\/ ($2.2 \cdot r_\mathrm{vdW}$) around
the closest hydrogen atom which are related to the van-der-Waals
radius of the hydrogen atom $r_\mathrm{vdW} =
1.2$\,\AA~\cite{BON:JPC64}. the MDCM reproduces the ESP grid points
within the range with a root mean squared error (RMSE) of
$0.73$\,kcal/mol per grid point.  In comparison, the RMSE within the
same range for a point charge model with charges set to zero is
$2.31$\,kcal/mol.  Contour plots of the reference ESP
($V_\mathrm{ref}$) on grid points at distances larger $1.44$\,\AA\/
from the closest hydrogen atom and the deviation from the model ESP
($V_\mathrm{MDCM}$) with $\Delta\mathrm{ESP} = V_\mathrm{MDCM} -
V_\mathrm{ref}$ are shown in Figure \ref{fig_esp_h2}A and B,
respectively.\\

\noindent
The computed components of the molecular quadrupole tensor
$\boldsymbol{Q}$ of H$_2$ are $\{ Q_{xx}, Q_{yy}, Q_{zz}\} = \{ -0.22,
-0.22, 0.45\}$\,D\AA\/, close to the measured experimental results of
$\{-0.26, -0.26, 0.52\}$\,D\AA~\cite{orcutt:1963}. The quadrupole
moment prediction of the fitted MDCM model for H$_2$ yields $\{0, 0,
0.61\}$\,D\AA. Both quadrupole moments $Q_{xx}$ and $Q_{yy}$ are zero
as the distributed charges are only displaced along the H$_2$ bond
axis but $Q_{zz}$ is about 35\% larger than the reference one but
close to the experimental result.\\

\section{Results}
First, the interaction between H$_2$ and the heme unit is
considered. Next, the structural dynamics and vibrational spectroscopy
of H$_2$ in the protein are analyzed from MD simulations.

\subsection{H$_2$ Interaction with Heme}
\begin{figure}[h]
\begin{center}
\includegraphics[width=0.95\textwidth]{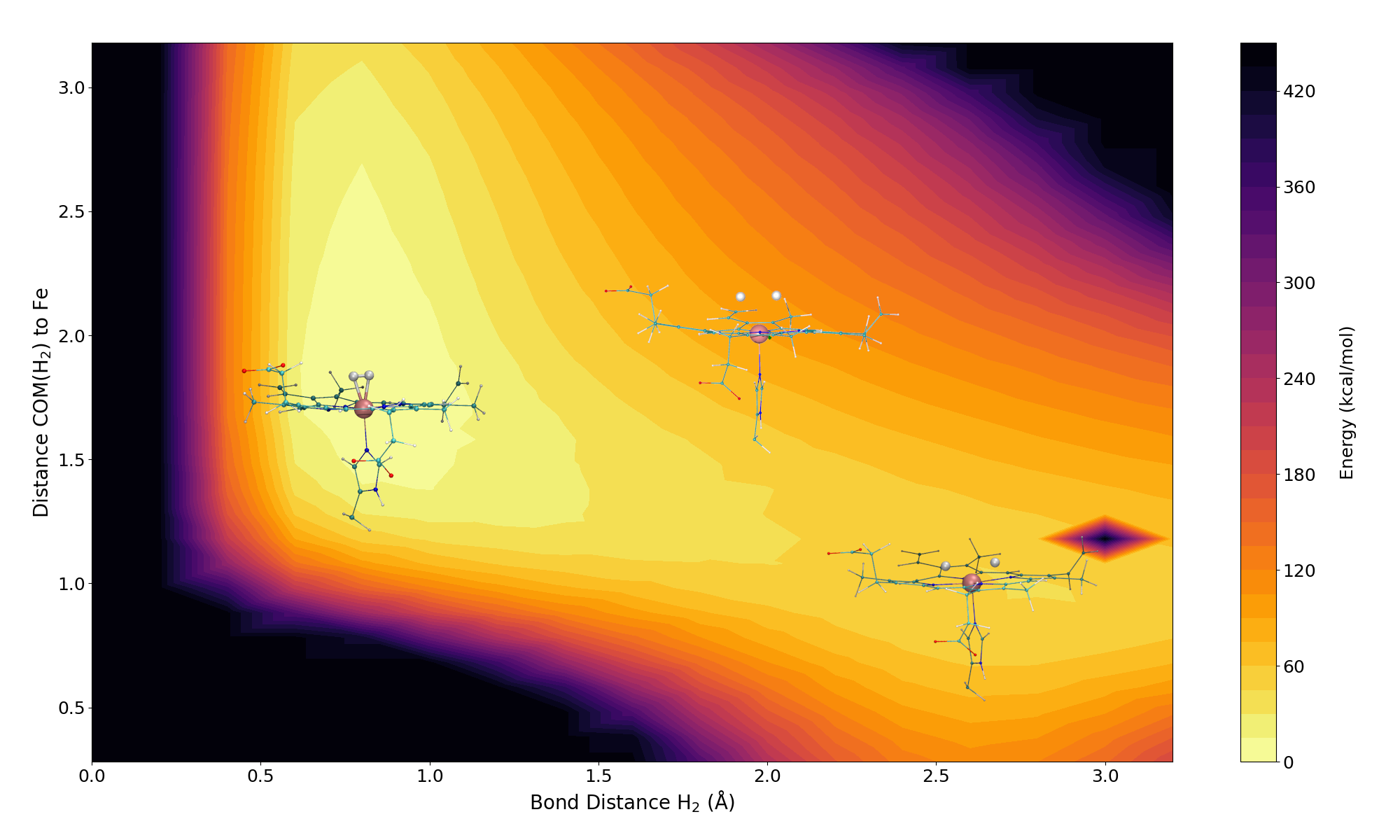}
\caption{{\bf PES scan of H$_2$ interacting with the Heme-Histidine
    active site of myoglobin.}  The coordinate system for scanning
  this potential energy surface is shown in Figure S1.}
\label{fig_pes}
\end{center}
\end{figure}

\begin{center}
\begin{figure}
\includegraphics[width=0.6\linewidth]{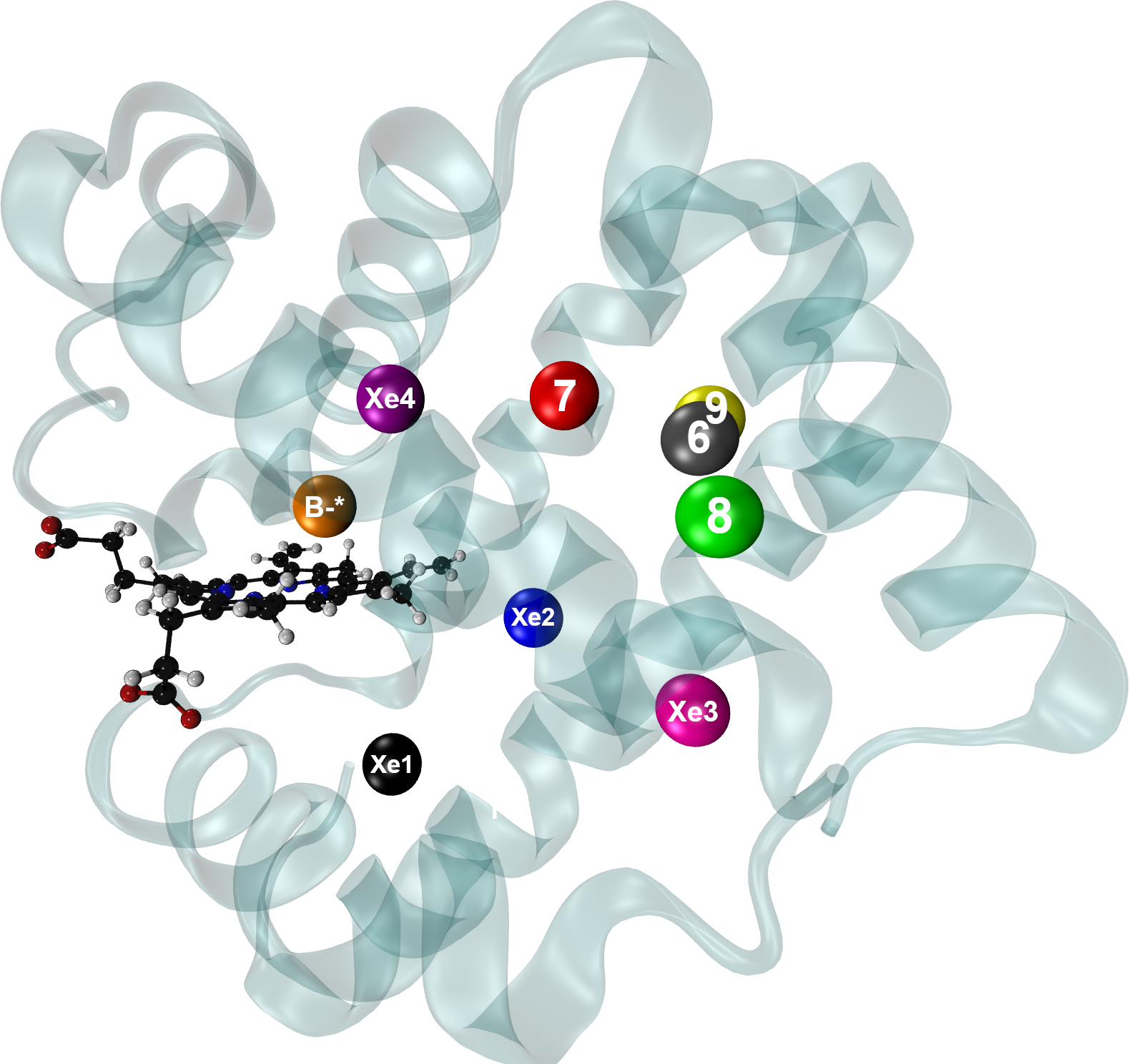}
\caption{{\bf Pocket representation in Mb.}  Shown is the secondary
  structure of Mb (8 helices) with the heme-unit in ball-and-stick
  representation together with the pockets determined for H$_2$ found
  in the present simulations. The pockets are Xe1 to Xe4, B-state and
  pockets 6 to 9 which were found in addition to the experimentally
  known ligand-binding sites \cite{tilton:1984,Lim:1995}.}
\label{fig_pocket_positions}
\end{figure}
\end{center}

The interaction between H$_2$ and the heme unit is shown in Figure
\ref{fig_pes}. With respect to H$_2$ outside the protein, the Fe-H$_2$
bound state is stabilized by $\sim -11.5$ kcal/mol. This is
considerably weaker than the interaction between heme and
physiologically relevant ligands -O$_2$ and -NO and poisonous -CO and
-CN$^-$. For -O$_2$ and -CO the experimentally determined standard
enthalpies of formation with myoglobin are $-18.1 \pm 0.4$ kcal/mol
and $-21.4 \pm 0.3$ kcal/mol, respectively \cite{lumry:1971}. No
direct experimental measurements exist for -NO and -CN$^-$ but density
functional theory calculations indicate comparable or stronger
interactions depending on the Fe-oxidation state: $-23.0$ kcal/mol for
Fe(II)-NO, more than $-33.0$ kcal/mol for Fe(III)-NO and stronger than
$-50.0$ kcal/mol for binding of -CN$^-$
\cite{MM.mbno:2005,MM.mbcn:2007}.\\

\noindent
The capability of the heme-Fe to break the H$_2$ bond was also
investigated. The H-Fe-H arrangement was found to be a faint minimum,
$13.9$\,kcal/mol above the dissociation limit or $25.4$\,kcal/mol
above the global minimum. The transition state separating the two
minima is $52.6$\,kcal/mol above the Fe-H$_2$ state,
i.e. $41.1$\,kcal/mol above the dissociation limit. Hence, H$_2$ binds
reversibly to heme-iron and no ``chemistry'' is expected to take
place.\\

\subsection{Structural Dynamics and H$_2$ Diffusion}
Next, the diffusional dynamics of H$_2$ within Mb was considered. MD
simulations were carried out using two energy functions. The first was
the conventional CGenFF energy function and for the second energy
function the H$_2$ molecule was described as a Morse oscillator and
MDCM electrostatics.  \\

\noindent
Using the CGenFF setup and MD simulation with five H$_2$ in Mb, it was
found that H$_2$ can localize in at least 9 different locations within
Mb. The first 4 pockets (Xe1 to Xe4) are the those found for xenon in
myoglobin \cite{tilton:1984,MM.mb:2023} together with the B-state
which was spectroscopically characterized
\cite{Lim:1995,schotte:2003,MM.mbco:2003,MM.mbco:2008,olson:2007}. Pockets
6 to 9 were detected from extended MD simulations. From visual
comparison, the newly determined were also observed as CO cavities in
previous publications \cite{bossa:2005,cohen:2006}, see e.g. Figure 1
in Ref \cite{bossa:2005}.  \\

\begin{figure}[htb!]
\centering
\includegraphics[width=0.95\linewidth]{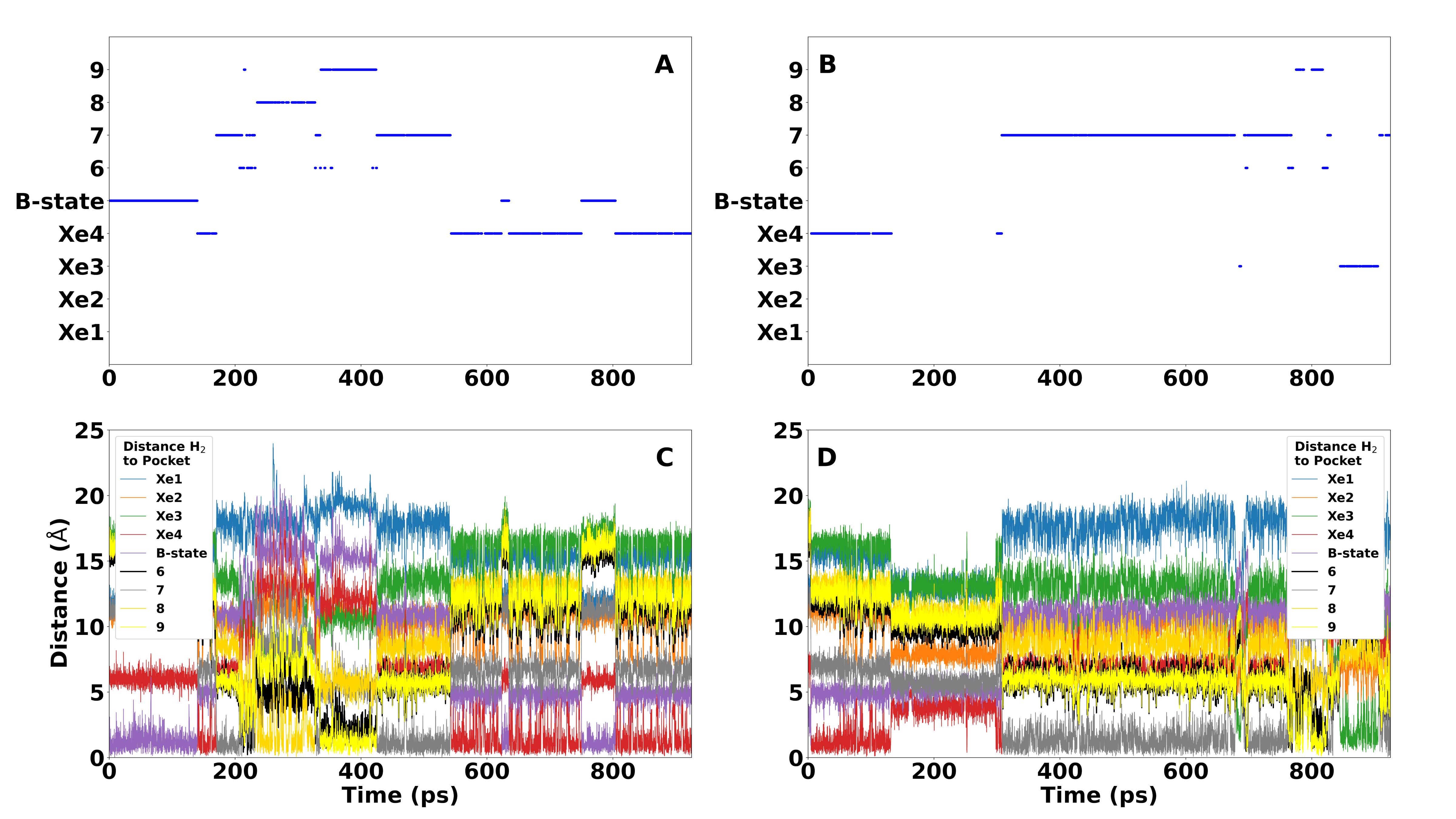}
\caption{{\bf Pocket dynamics of H$_2$ in Mb.}  Panels A and B report
  the pocket occupied by H$_2$ as a function of simulation time from
  simulations using the CGenFF and MDCM/Morse energy functions,
  respectively. Panels C and D show the separation between H$_2$ and
  each of the pocket centers (Xe1 to Xe4, B-state and 6 to 9). Each
  color corresponds to a particular separation between H$_2$ and the
  respective pocket center. In panel D between 150 and 300 ps the
  distance between H$_2$ and any other pocket is $\sim 5$ \AA\/ which
  points to one or several other uncharacterized docking sites.}
\label{fig_H2Diffusion}
\end{figure}

\noindent
Time series for the separation of the center of mass of H$_2$ to each
of the nine pocket centers are reported in Figures
\ref{fig_H2Diffusion}C and D. The pocket centers were determined from
the procedure described in the Methods section applied to MD
simulations with one H$_2$ in Mb. It is found that the H$_2$ molecule
readily migrates between the different pockets (see traces for
distances $\sim 2$ \AA). For example, the simulation using the CGenFF
energy function finds H$_2$ visiting 6 out of the 9 pockets during
600\,ps. Similarly using the MDCM model for the electrostatic (panel
D) also leads to diffusion but only 4 different pockets are visited
within the same time frame. This indicates that the interaction
between H$_2$ and the protein environment is stronger due to the
modified electrostatics. The unassigned spatial location between 150
and 300 ps is likely to be a sub-pocket of Xe4 which was also found
for CO diffusing through Mb (called Xe4(2)).\cite{MM.rough:2007,
  MM.mbco:2005}\\

\subsection{H$_2$ Vibrational Spectra}
Next, the vibrational spectroscopy of H$_2$ within the protein was
analyzed by computing the power spectra from the H$_2$ bond
distance. It is of interest to assess whether the electrostatic
interaction between the protein environment and H$_2$ leads to
pocket-specific spectra once MDCM as the electrostatic model for the
diatomic is used. Two types of simulations, unconstrained and
constrained, were carried out.  The unconstrained MD simulations were
initialized with H$_2$ close to the heme-Fe, above the porphyrin plane
in the distal site of Mb (B-state). Second, to investigate the impact
of the pocket positions on the vibrational spectroscopy of H$_2$,
constrained simulations were performed in which H$_2$ was weakly
harmonically constrained to each center of mass of one of the nine
pockets to obtain pocket-specific spectra.\\

\begin{figure}
\centering
\includegraphics[width=0.95\linewidth]{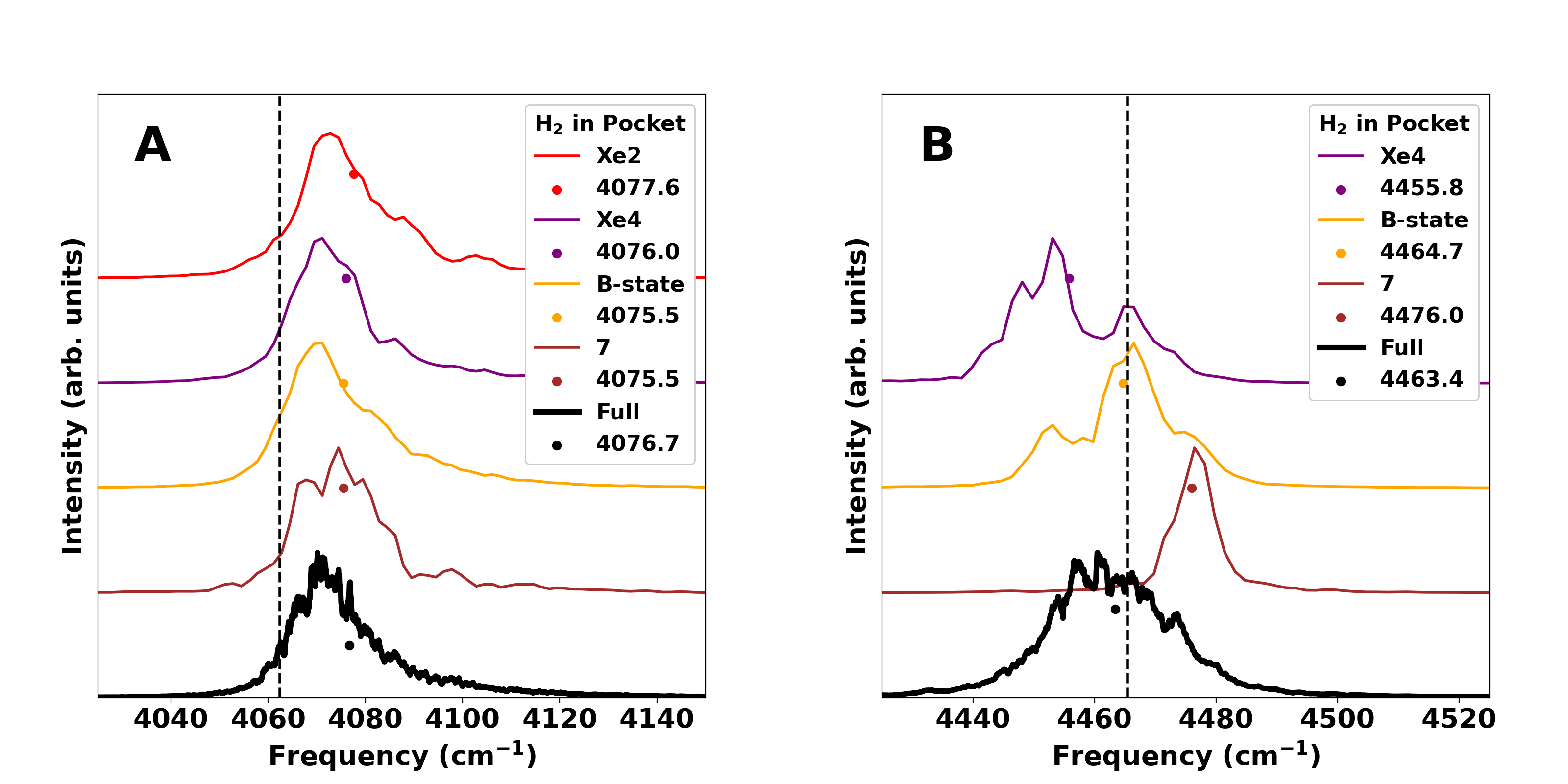}
\caption{{\bf Vibrational spectra of H$_2$ in Mb from unconstrained MD
    simulations.}  Panel A: simulations using CGenFF and panel B:
  simulations using the Morse-potential and MDCM for H$_2$. The black
  trace is the total spectrum as it would, for example, be measured
  from an experiment.  Filled circles indicate the mean of each
  spectrum and the mean frequencies are given in the legend and the
  simulated frequency of gas phase H$_2$ is shown as vertical dashed
  line at (A) $4062.4$ cm$^{-1}$ and (B) $4465.4$ cm$^{-1}$.}
\label{fig_spec1}
\end{figure}

\noindent
Figure \ref{fig_spec1} reports the vibrational spectrum from the free
dynamics which featured a single H$_2$ in Mb.  Only the times during
where H$_2$ was within one of the specified pockets were
analyzed. Because the residence times of the ligand in each of the
pockets differs, sampling times also differ. In Figure
\ref{fig_spec1}A (CGenFF with harmonic H$_2$), the power spectra are
all broad and feature a single maximum, except for that associated
with pocket 7. There is a slight shift of the mean peak positions of
each spectrum with that of pocket Xe2 most shifted to the blue by
$+16.2$ cm$^{-1}$ (maximum at $4078.6$ cm$^{-1}$) and that of pocket 7
least shifted to the blue by $+9.1$ cm$^{-1}$ (at $4071.5$ cm$^{-1}$),
away from the harmonic gas phase spectrum. The blue shift also
indicates that the H$_2$ bond is slightly strengthened, due to
repulsive van der Waals interactions between the unbound ligand and
the surrounding protein.\\

\noindent
For simulations using the Morse potential for the H$_2$ bond and the
MDCM model to include the molecular quadrupole moment, the power
spectra are reported in Figure \ref{fig_spec1}B. Here, the ligand
samples pockets Xe4, B-state, and 7 but not pocket Xe2. With the
refined energy function the spectra differ in their widths both
between the pockets and compared with simulations using the CGenFF
energy function (see panel A). For example, the spectrum associated
with pocket 7 is rather narrow whereas that for pocket Xe4 is overall
broad but consists of two separate peaks. Also, the pocket-specific
spectra feature both, red- and blue-shifts which indicate slight
weakening and strengthening of the H$_2$ bond due to favourable and
unfavourable intermolecular interactions with the protein
environment. For H$_2$ in pocket Xe4 the red shift with respect to the
gas phase spectrum amounts to $-9.3$ cm$^{-1}$ whereas the blue shift
for pocket 7 is $+10.8$ cm$^{-1}$.\\

\begin{figure}
\includegraphics[width=0.95\linewidth]{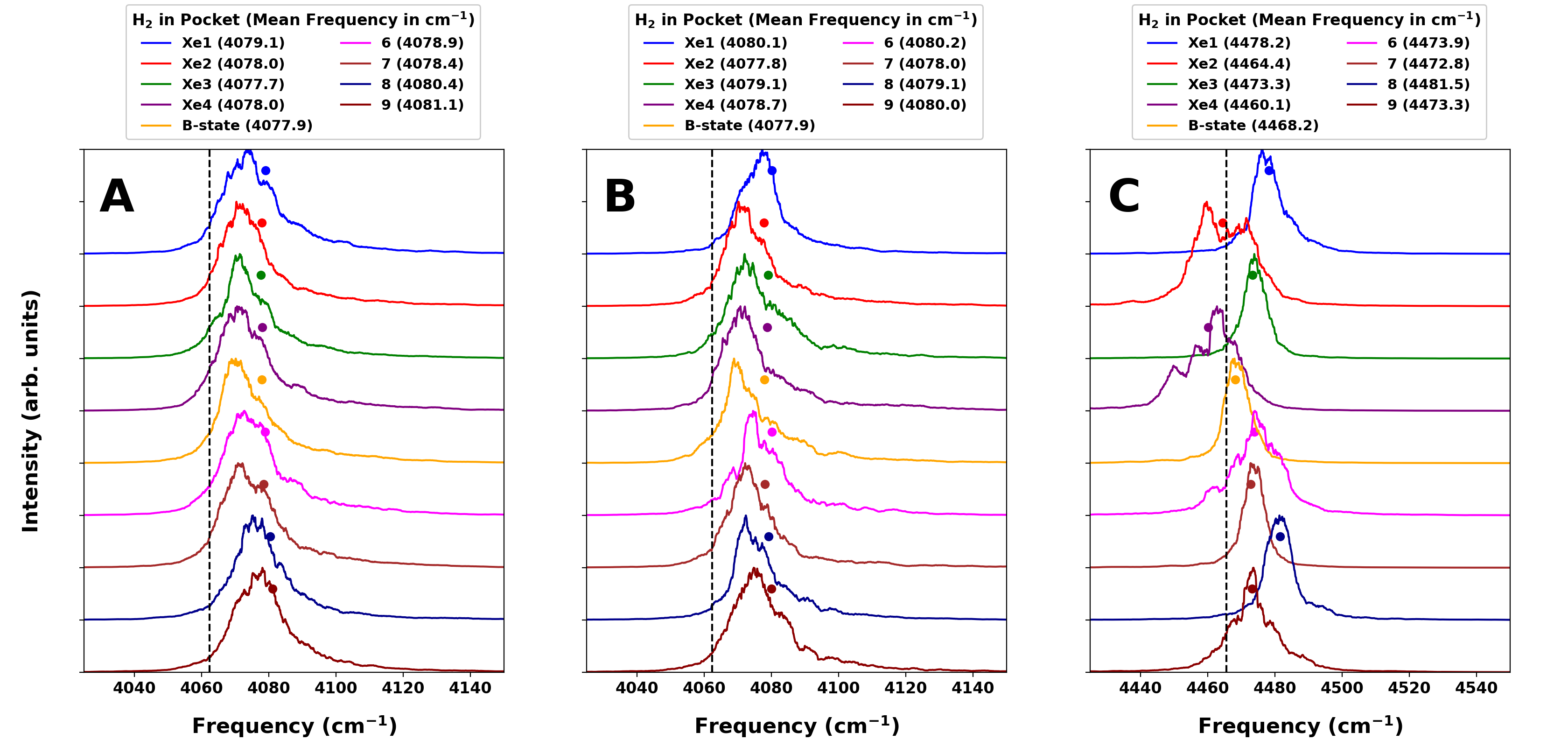}
\caption{{\bf Vibrational spectra of H$_2$ in Mb from
    pocket-constrained MD simulations.}  Panel A: simulations using
  the CGenFF energy function, Panel B: using the MDCM model for H$_2$
  but a conventional harmonic bond potential and Panel C: using the
  MDCM model and the Morse potential for H$_2$. The weighted average
  position of the maximum intensity (in cm$^{-1}$) for each spectra is
  given in brackets in the legend. The vibrational frequency from MD
  simulation of H$_2$ in the gas phase is marked as a vertical dashed
  line.}
\label{fig_spec2}
\end{figure}

\noindent
Next, the pocket specific spectra are analyzed, see Figure
\ref{fig_spec2}. In all simulations, the H$_2$ ligand was slightly
constrained towards the center of each of the 9 pockets using a
harmonic constraint. This was to ensure that pocket-specific spectra
can be obtained. Three types of simulations were carried out: one
using the CGenFF force field, a second one using the MDCM charge model
but the harmonic bond potential and the third combining MDCM with the
Morse-bond for H$_2$.\\

\noindent
Broadly speaking, the power spectra for H$_2$ in all nine pockets from
simulations using the CGenFF force field are rather similar, not to
say largely identical, see Figure \ref{fig_spec2}A. All peaks are
shifted to the blue away from the gas phase spectrum and the peak
positions cover a range of only 3 cm$^{-1}$ (from 4078 cm$^{-1}$ to
4081 cm$^{-1}$). Using the harmonic bond together with the MDCM model
leads to similar observations, as shown in Figure
\ref{fig_spec2}B. Including, however, a Morse-description for the
H$_2$ bond leads to considerable changes, see \ref{fig_spec2}C. Now
both, blue- and red-shifts of the power spectra appear and the maxima
of the spectra cover a considerably wider range, covering 22 cm$^{-1}$
(4460 cm$^{-1}$ to 4482 cm$^{-1}$). This finding suggests that only
the combination of an improved description of the bonded interaction
(Morse) together with modeling the electrostatics (quadrupole from
MDCM) provides the necessary detail to yield the expected response of
H$_2$ to the inhomogeneous electric field in each of the protein
pockets as was also observed for CO from both, simulations and
experiments
\cite{Lim:1995,Lim:1997,MM.mbco:2003,Merchant03,MM.mbco:2004,MM.mbco:2006,MM.mbco:2008}.\\

\section{Discussion and Conclusions}
In this work the interaction of H$_2$ with the heme-group of Mb and
the diffusional dynamics of the diatomic within the protein was
investigated.  This was motivated by the observation that over the
past 15 years H$_2$ has emerged as a physiologically interesting
``small molecule'' akin to the well known diatomics O$_2$ or NO. The
results of the present study demonstrate that H$_2$ can reside within
Mb, occupies the same spatial regions as those found experimentally
and from computations for Xe, CO, and NO, and can even populate less
well-characterized internal defects.\\

\noindent
The magnitude of the quadrupole moment of H$_2$ is approximately half
that of carbon monoxide (CO) for which a value of $\Theta = -9.47
\times 10^{-40}$ Cm$^2$ was found experimentally \cite{raab:1998},
compared with calculations of $\Theta \sim -2$ D\AA\/, corresponding
to $\sim -6.7 \times 10^{-40}$ Cm$^2$ \cite{maroulis:2001}. Hence, the
Stark shifts originating from the interaction between the electrical
moments of the ligand and the electrical field of the surrounding
protein are expected to be comparable, but somewhat smaller for H$_2$
compared with CO
\cite{Lim:1995,Lim:1997,MM.mbco:2003,MM.mbco:2004,MM.mbco:2008}. For
CO in Mb the Stark shift of the split IR spectrum is --10 cm$^{-1}$
and --20 cm$^{-1}$ for the two experimentally measured bands which
correspond to Fe--CO and Fe--OC orientations in the
B-state.\cite{Lim:1995,MM.mbco:2008,MM.mbco:2003,MM.mbco:2004} For the
B-state the frequency shift for H$_2$ is $\sim 4$ cm$^{-1}$ to the
blue, see Figure \ref{fig_spec2}C but for other pockets the shift can
be considerably larger. Hence, it might also be of interest to further
characterize the pocket-specific infrared spectra for CO in pockets
other than the B-state and Xe4.\\

\noindent
Comparing the pocket specific vibrational spectra it is found that the
H$_2$ vibrational spectra change as the ligand occupies different
regions within the protein, see Figure \ref{fig_spec2}. Because H$_2$
has a vanishing dipole moment, it is more likely to observe its Raman
spectrum \cite{teal:1935}. The use of Raman spectroscopy to query
myoglobin is well established \cite{hu:1996} and the frequency range
for H$_2$ above 4000 cm$^{-1}$ is well removed from other
spectroscopic signatures of the protein.\\

\noindent
In view of chemical reactivity of H$_2$ towards heme, Figure
\ref{fig_pes} establishes the existence of two low-energy states. The
first is located for which the separation between H$_2$ center of mass
and Fe is 1.5 \AA\/ and the distance between the H atoms is 0.81
\AA. For the second low energy state the distance between the H$_2$
center of mass and Fe is $\sim 1$ \AA\/ but with a direct Fe-H
distance of 1.8 \AA.  The first state was expected as the distance
between the H two atoms is comparable to the equilibrium separation
for H$_2$ in the gas phase. The second state is more interesting since
the H atoms are separated by $\sim 2.7$ \AA\/ which suggests that the
H--H bond can be broken when bound to Fe, and remain
stable. Nevertheless, reaching the H-Fe-H dissociated state involves a
high barrier ($\sim 50$ kcal/mol) and is physiologically irrelevant.\\

\noindent
In conclusion the present work finds that the vibrational spectroscopy
of H$_2$ is sensitive to the chemical environment. H$_2$ as a ligand
is well-tolerated within myoglobin and given the role of H$_2$ for
various physiological processes it is of interest to further
characterize the interaction between H$_2$ and myoglobin from an
experimental perspective. It is hoped that the present work provides
an initial stimulus for such studies.\\

\section*{Acknowledgment}
This work has been financially supported by the Swiss National Science
Foundation (NCCR-MUST, grants 200021-117810, 200020-188724), the
University of Basel and by the European Union's Horizon 2020 research
and innovation program under the Marie Sk{\l}odowska-Curie grant
agreement No 801459 -FP-RESOMUS.\\

\section*{Supporting Information}
The supporting material reports the residues defining pockets Xe1 to
Xe4, B-state and pockets 6 to 9 and the coordinate system used for
scanning the PES for H$_2$ interacting with heme.

\section*{Data Availability}
Relevant source data and evaluation files for the present results are
available at \url{https://github.com/MMunibas/H2-Myoglobin}.

\section*{Graphical Abstract}
\includegraphics[width=0.95\linewidth]{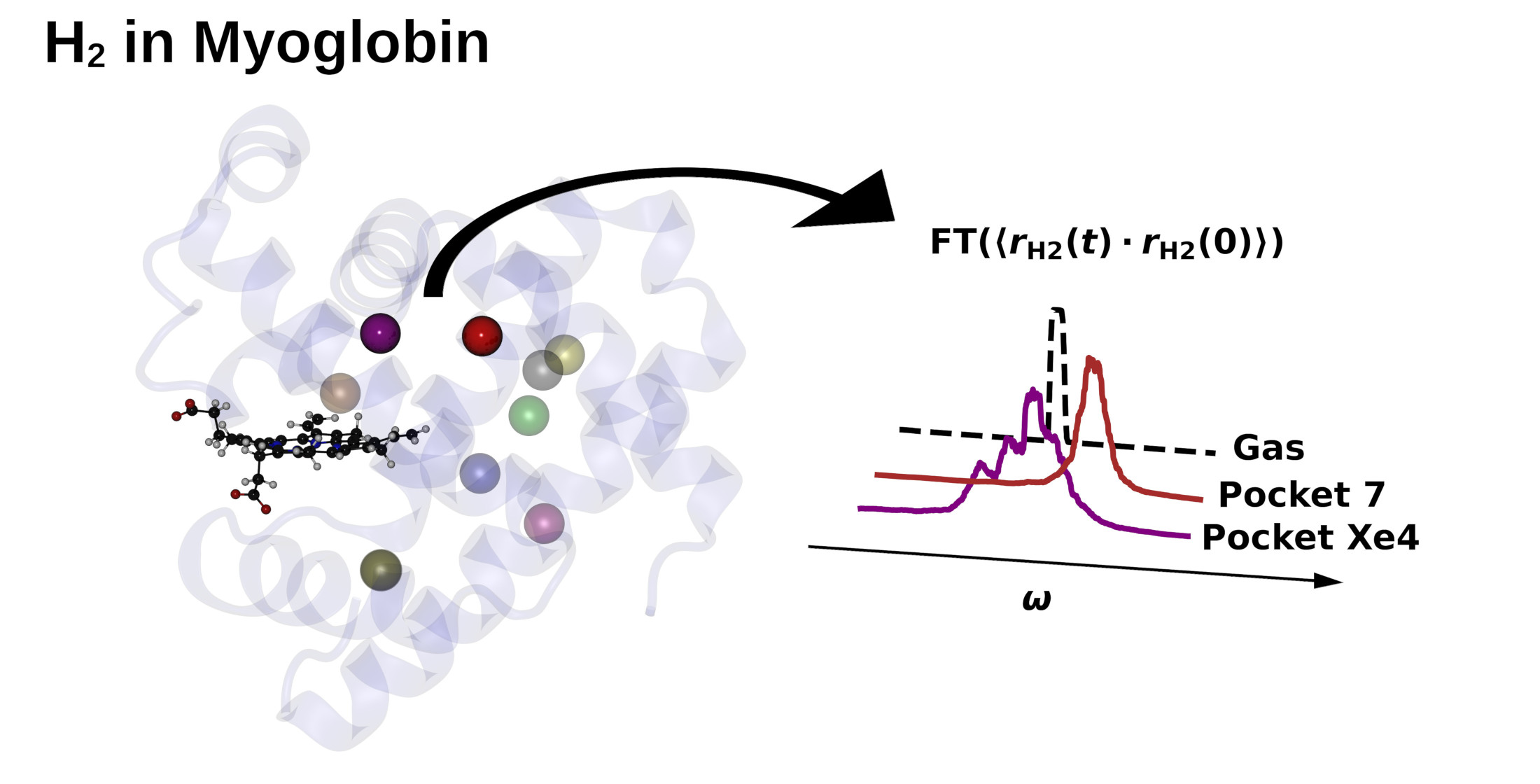}

\bibliography{refs}

\providecommand{\latin}[1]{#1}
\makeatletter
\providecommand{\doi}
  {\begingroup\let\do\@makeother\dospecials
  \catcode`\{=1 \catcode`\}=2 \doi@aux}
\providecommand{\doi@aux}[1]{\endgroup\texttt{#1}}
\makeatother
\providecommand*\mcitethebibliography{\thebibliography}
\csname @ifundefined\endcsname{endmcitethebibliography}
  {\let\endmcitethebibliography\endthebibliography}{}
\begin{mcitethebibliography}{65}
\providecommand*\natexlab[1]{#1}
\providecommand*\mciteSetBstSublistMode[1]{}
\providecommand*\mciteSetBstMaxWidthForm[2]{}
\providecommand*\mciteBstWouldAddEndPuncttrue
  {\def\EndOfBibitem{\unskip.}}
\providecommand*\mciteBstWouldAddEndPunctfalse
  {\let\EndOfBibitem\relax}
\providecommand*\mciteSetBstMidEndSepPunct[3]{}
\providecommand*\mciteSetBstSublistLabelBeginEnd[3]{}
\providecommand*\EndOfBibitem{}
\mciteSetBstSublistMode{f}
\mciteSetBstMaxWidthForm{subitem}{(\alph{mcitesubitemcount})}
\mciteSetBstSublistLabelBeginEnd
  {\mcitemaxwidthsubitemform\space}
  {\relax}
  {\relax}

\bibitem[Austin \latin{et~al.}(1975)Austin, Beeson, Eisenstein, Frauenfelder,
  and Gunsalus]{austin:1975}
Austin,~R.~H.; Beeson,~K.; Eisenstein,~L.; Frauenfelder,~H.; Gunsalus,~I.
  Dynamics of ligand binding to myoglobin. \emph{Biochem.} \textbf{1975},
  \emph{14}, 5355--5373\relax
\mciteBstWouldAddEndPuncttrue
\mciteSetBstMidEndSepPunct{\mcitedefaultmidpunct}
{\mcitedefaultendpunct}{\mcitedefaultseppunct}\relax
\EndOfBibitem
\bibitem[Cornelius \latin{et~al.}(1983)Cornelius, Hochstrasser, and
  Steele]{cornelius:1983}
Cornelius,~P.; Hochstrasser,~R.; Steele,~A. Ultrafast relaxation in picosecond
  photolysis of nitrosylhemoglobin. \emph{J. Mol. Biol.} \textbf{1983},
  \emph{163}, 119--128\relax
\mciteBstWouldAddEndPuncttrue
\mciteSetBstMidEndSepPunct{\mcitedefaultmidpunct}
{\mcitedefaultendpunct}{\mcitedefaultseppunct}\relax
\EndOfBibitem
\bibitem[Martin \latin{et~al.}(1983)Martin, Migus, Poyart, Lecarpentier,
  Astier, and Antonetti]{martin:1983}
Martin,~J.; Migus,~A.; Poyart,~C.; Lecarpentier,~Y.; Astier,~R.; Antonetti,~A.
  Femtosecond photolysis of CO-ligated protoheme and hemoproteins: appearance
  of deoxy species with a 350-fsec time constant. \emph{Proc. Natl. Acad. Sci.}
  \textbf{1983}, \emph{80}, 173--177\relax
\mciteBstWouldAddEndPuncttrue
\mciteSetBstMidEndSepPunct{\mcitedefaultmidpunct}
{\mcitedefaultendpunct}{\mcitedefaultseppunct}\relax
\EndOfBibitem
\bibitem[Steinbach \latin{et~al.}(1991)Steinbach, Ansari, Berendzen,
  Braunstein, Chu, Cowen, Ehrenstein, Frauenfelder, and
  Johnson]{steinbach:1991}
Steinbach,~P.~J.; Ansari,~A.; Berendzen,~J.; Braunstein,~D.; Chu,~K.;
  Cowen,~B.~R.; Ehrenstein,~D.; Frauenfelder,~H.; Johnson,~J.~B. Ligand binding
  to heme proteins: connection between dynamics and function. \emph{Biochem.}
  \textbf{1991}, \emph{30}, 3988--4001\relax
\mciteBstWouldAddEndPuncttrue
\mciteSetBstMidEndSepPunct{\mcitedefaultmidpunct}
{\mcitedefaultendpunct}{\mcitedefaultseppunct}\relax
\EndOfBibitem
\bibitem[Schotte \latin{et~al.}(2003)Schotte, Lim, Jackson, Smirnov, Soman,
  Olson, Phillips~Jr, Wulff, and Anfinrud]{schotte:2003}
Schotte,~F.; Lim,~M.; Jackson,~T.~A.; Smirnov,~A.~V.; Soman,~J.; Olson,~J.~S.;
  Phillips~Jr,~G.~N.; Wulff,~M.; Anfinrud,~P.~A. Watching a protein as it
  functions with 150-ps time-resolved x-ray crystallography. \emph{Science}
  \textbf{2003}, \emph{300}, 1944--1947\relax
\mciteBstWouldAddEndPuncttrue
\mciteSetBstMidEndSepPunct{\mcitedefaultmidpunct}
{\mcitedefaultendpunct}{\mcitedefaultseppunct}\relax
\EndOfBibitem
\bibitem[Barends \latin{et~al.}(2015)Barends, Foucar, Ardevol, Nass, Aquila,
  Botha, Doak, Falahati, Hartmann, Hilpert, \latin{et~al.}
  others]{barends:2015}
Barends,~T.~R.; Foucar,~L.; Ardevol,~A.; Nass,~K.; Aquila,~A.; Botha,~S.;
  Doak,~R.~B.; Falahati,~K.; Hartmann,~E.; Hilpert,~M. \latin{et~al.}  Direct
  observation of ultrafast collective motions in CO myoglobin upon ligand
  dissociation. \emph{Science} \textbf{2015}, \emph{350}, 445--450\relax
\mciteBstWouldAddEndPuncttrue
\mciteSetBstMidEndSepPunct{\mcitedefaultmidpunct}
{\mcitedefaultendpunct}{\mcitedefaultseppunct}\relax
\EndOfBibitem
\bibitem[Barends \latin{et~al.}(2024)Barends, Gorel, Bhattacharyya, Schir{\`o},
  Bacellar, Cirelli, Colletier, Foucar, Gr{\"u}nbein, Hartmann, \latin{et~al.}
  others]{barends:2024}
Barends,~T.~R.; Gorel,~A.; Bhattacharyya,~S.; Schir{\`o},~G.; Bacellar,~C.;
  Cirelli,~C.; Colletier,~J.-P.; Foucar,~L.; Gr{\"u}nbein,~M.~L.; Hartmann,~E.
  \latin{et~al.}  Influence of pump laser fluence on ultrafast myoglobin
  structural dynamics. \emph{Nature} \textbf{2024}, \emph{626}, 905--911\relax
\mciteBstWouldAddEndPuncttrue
\mciteSetBstMidEndSepPunct{\mcitedefaultmidpunct}
{\mcitedefaultendpunct}{\mcitedefaultseppunct}\relax
\EndOfBibitem
\bibitem[Nutt and Meuwly(2004)Nutt, and Meuwly]{MM.mbco:2004}
Nutt,~D.~R.; Meuwly,~M. CO migration in native and mutant myoglobin: atomistic
  simulations for the understanding of protein function. \emph{Proc. Natl.
  Acad. Sci.} \textbf{2004}, \emph{101}, 5998--6002\relax
\mciteBstWouldAddEndPuncttrue
\mciteSetBstMidEndSepPunct{\mcitedefaultmidpunct}
{\mcitedefaultendpunct}{\mcitedefaultseppunct}\relax
\EndOfBibitem
\bibitem[Meuwly \latin{et~al.}(2002)Meuwly, Becker, Stote, and
  Karplus]{MM.mbno:2002}
Meuwly,~M.; Becker,~O.~M.; Stote,~R.; Karplus,~M. NO rebinding to myoglobin: a
  reactive molecular dynamics study. \emph{Biophys. Chem.} \textbf{2002},
  \emph{98}, 183--207\relax
\mciteBstWouldAddEndPuncttrue
\mciteSetBstMidEndSepPunct{\mcitedefaultmidpunct}
{\mcitedefaultendpunct}{\mcitedefaultseppunct}\relax
\EndOfBibitem
\bibitem[Soloviov and Meuwly(2015)Soloviov, and Meuwly]{MM.mbno:2015}
Soloviov,~M.; Meuwly,~M. Reproducing kernel potential energy surfaces in
  biomolecular simulations: Nitric oxide binding to myoglobin. \emph{J. Chem.
  Phys.} \textbf{2015}, \emph{143}\relax
\mciteBstWouldAddEndPuncttrue
\mciteSetBstMidEndSepPunct{\mcitedefaultmidpunct}
{\mcitedefaultendpunct}{\mcitedefaultseppunct}\relax
\EndOfBibitem
\bibitem[Soloviov \latin{et~al.}(2016)Soloviov, Das, and Meuwly]{MM.mbno:2016}
Soloviov,~M.; Das,~A.~K.; Meuwly,~M. Structural Interpretation of Metastable
  States in Myoglobin--NO. \emph{Angew. Chem. Int. Ed.} \textbf{2016},
  \emph{55}, 10126--10130\relax
\mciteBstWouldAddEndPuncttrue
\mciteSetBstMidEndSepPunct{\mcitedefaultmidpunct}
{\mcitedefaultendpunct}{\mcitedefaultseppunct}\relax
\EndOfBibitem
\bibitem[Cohen \latin{et~al.}(2006)Cohen, Arkhipov, Braun, and
  Schulten]{cohen:2006}
Cohen,~J.; Arkhipov,~A.; Braun,~R.; Schulten,~K. Imaging the migration pathways
  for O$_2$, CO, NO, and Xe inside myoglobin. \emph{Biophys. J.} \textbf{2006},
  \emph{91}, 1844--1857\relax
\mciteBstWouldAddEndPuncttrue
\mciteSetBstMidEndSepPunct{\mcitedefaultmidpunct}
{\mcitedefaultendpunct}{\mcitedefaultseppunct}\relax
\EndOfBibitem
\bibitem[Tilton~Jr \latin{et~al.}(1984)Tilton~Jr, Kuntz~Jr, and
  Petsko]{tilton:1984}
Tilton~Jr,~R.~F.; Kuntz~Jr,~I.~D.; Petsko,~G.~A. Cavities in proteins:
  structure of a metmyoglobin xenon complex solved to 1.9. ANG. \emph{Biochem.}
  \textbf{1984}, \emph{23}, 2849--2857\relax
\mciteBstWouldAddEndPuncttrue
\mciteSetBstMidEndSepPunct{\mcitedefaultmidpunct}
{\mcitedefaultendpunct}{\mcitedefaultseppunct}\relax
\EndOfBibitem
\bibitem[Olson \latin{et~al.}(2007)Olson, Soman, and Phillips~Jr]{olson:2007}
Olson,~J.~S.; Soman,~J.; Phillips~Jr,~G. Ligand pathways in myoglobin: a review
  of Trp cavity mutations. \emph{IUBMB life} \textbf{2007}, \emph{59},
  552--562\relax
\mciteBstWouldAddEndPuncttrue
\mciteSetBstMidEndSepPunct{\mcitedefaultmidpunct}
{\mcitedefaultendpunct}{\mcitedefaultseppunct}\relax
\EndOfBibitem
\bibitem[Turan \latin{et~al.}(2023)Turan, Boittier, and Meuwly]{MM.mb:2023}
Turan,~H.~T.; Boittier,~E.; Meuwly,~M. Interaction at a distance: {Xenon}
  migration in {Mb}. \emph{J. Chem. Phys.} \textbf{2023}, \emph{158},
  125103\relax
\mciteBstWouldAddEndPuncttrue
\mciteSetBstMidEndSepPunct{\mcitedefaultmidpunct}
{\mcitedefaultendpunct}{\mcitedefaultseppunct}\relax
\EndOfBibitem
\bibitem[Milani \latin{et~al.}(2004)Milani, Pesce, Ouellet, Dewilde, Friedman,
  Ascenzi, Guertin, and Bolognesi]{milani:2004}
Milani,~M.; Pesce,~A.; Ouellet,~Y.; Dewilde,~S.; Friedman,~J.; Ascenzi,~P.;
  Guertin,~M.; Bolognesi,~M. Heme-ligand tunneling in group I truncated
  hemoglobins. \emph{J. Biol. Chem.} \textbf{2004}, \emph{279},
  21520--21525\relax
\mciteBstWouldAddEndPuncttrue
\mciteSetBstMidEndSepPunct{\mcitedefaultmidpunct}
{\mcitedefaultendpunct}{\mcitedefaultseppunct}\relax
\EndOfBibitem
\bibitem[Cazade and Meuwly(2012)Cazade, and Meuwly]{MM.trhb:2012}
Cazade,~P.-A.; Meuwly,~M. Oxygen Migration Pathways in NO-bound Truncated
  Hemoglobin. \emph{Chem. Phys. Chem.} \textbf{2012}, \emph{13},
  4276--4286\relax
\mciteBstWouldAddEndPuncttrue
\mciteSetBstMidEndSepPunct{\mcitedefaultmidpunct}
{\mcitedefaultendpunct}{\mcitedefaultseppunct}\relax
\EndOfBibitem
\bibitem[Brunori and Vallone(2007)Brunori, and Vallone]{brunori:2007}
Brunori,~M.; Vallone,~B. Neuroglobin, seven years after. \emph{Cell. Mol. Life
  Sci.} \textbf{2007}, \emph{64}, 1259--1268\relax
\mciteBstWouldAddEndPuncttrue
\mciteSetBstMidEndSepPunct{\mcitedefaultmidpunct}
{\mcitedefaultendpunct}{\mcitedefaultseppunct}\relax
\EndOfBibitem
\bibitem[Lutz \latin{et~al.}(2009)Lutz, Nienhaus, Nienhaus, and
  Meuwly]{MM.ngb:2009}
Lutz,~S.; Nienhaus,~K.; Nienhaus,~G.~U.; Meuwly,~M. Ligand migration between
  internal docking sites in photodissociated carbonmonoxy neuroglobin. \emph{J.
  Phys. Chem. B} \textbf{2009}, \emph{113}, 15334--15343\relax
\mciteBstWouldAddEndPuncttrue
\mciteSetBstMidEndSepPunct{\mcitedefaultmidpunct}
{\mcitedefaultendpunct}{\mcitedefaultseppunct}\relax
\EndOfBibitem
\bibitem[Ohsawa \latin{et~al.}(2007)Ohsawa, Ishikawa, Takahashi, Watanabe,
  Nishimaki, Yamagata, Katsura, Katayama, Asoh, and Ohta]{ohsawa:2007}
Ohsawa,~I.; Ishikawa,~M.; Takahashi,~K.; Watanabe,~M.; Nishimaki,~K.;
  Yamagata,~K.; Katsura,~K.-i.; Katayama,~Y.; Asoh,~S.; Ohta,~S. Hydrogen acts
  as a therapeutic antioxidant by selectively reducing cytotoxic oxygen
  radicals. \emph{Nat. Med.} \textbf{2007}, \emph{13}, 688--694\relax
\mciteBstWouldAddEndPuncttrue
\mciteSetBstMidEndSepPunct{\mcitedefaultmidpunct}
{\mcitedefaultendpunct}{\mcitedefaultseppunct}\relax
\EndOfBibitem
\bibitem[Dole \latin{et~al.}(1975)Dole, Wilson, and Fife]{dole:1975}
Dole,~M.; Wilson,~F.~R.; Fife,~W.~P. Hyperbaric hydrogen therapy: a possible
  treatment for cancer. \emph{Science} \textbf{1975}, \emph{190},
  152--154\relax
\mciteBstWouldAddEndPuncttrue
\mciteSetBstMidEndSepPunct{\mcitedefaultmidpunct}
{\mcitedefaultendpunct}{\mcitedefaultseppunct}\relax
\EndOfBibitem
\bibitem[Wu \latin{et~al.}(2023)Wu, Zou, Feng, Zhu, Li, Liu, Duan, and
  Yang]{wu:2023}
Wu,~C.; Zou,~P.; Feng,~S.; Zhu,~L.; Li,~F.; Liu,~T. C.-Y.; Duan,~R.; Yang,~L.
  Molecular hydrogen: an emerging therapeutic medical gas for brain disorders.
  \emph{Mol. Neur.} \textbf{2023}, \emph{60}, 1749--1765\relax
\mciteBstWouldAddEndPuncttrue
\mciteSetBstMidEndSepPunct{\mcitedefaultmidpunct}
{\mcitedefaultendpunct}{\mcitedefaultseppunct}\relax
\EndOfBibitem
\bibitem[Fu \latin{et~al.}(2022)Fu, Zhang, and Zhang]{fu:2022}
Fu,~Z.; Zhang,~J.; Zhang,~Y. Role of Molecular Hydrogen in Ageing and
  Ageing-Related Diseases. \emph{Oxid. Med. Cell Longev.} \textbf{2022},
  \emph{2022}, 2249749\relax
\mciteBstWouldAddEndPuncttrue
\mciteSetBstMidEndSepPunct{\mcitedefaultmidpunct}
{\mcitedefaultendpunct}{\mcitedefaultseppunct}\relax
\EndOfBibitem
\bibitem[Szab{\'o}(2007)]{szabo:2007}
Szab{\'o},~C. Hydrogen sulphide and its therapeutic potential. \emph{Nat. Rev.
  Drug. Disc.} \textbf{2007}, \emph{6}, 917--935\relax
\mciteBstWouldAddEndPuncttrue
\mciteSetBstMidEndSepPunct{\mcitedefaultmidpunct}
{\mcitedefaultendpunct}{\mcitedefaultseppunct}\relax
\EndOfBibitem
\bibitem[Kajimura \latin{et~al.}(2010)Kajimura, Fukuda, Bateman, Yamamoto, and
  Suematsu]{kajimura:2010}
Kajimura,~M.; Fukuda,~R.; Bateman,~R.~M.; Yamamoto,~T.; Suematsu,~M.
  Interactions of multiple gas-transducing systems: hallmarks and uncertainties
  of CO, NO, and H$_2$S gas biology. \emph{Antiox. Red. Sig.} \textbf{2010},
  \emph{13}, 157--192\relax
\mciteBstWouldAddEndPuncttrue
\mciteSetBstMidEndSepPunct{\mcitedefaultmidpunct}
{\mcitedefaultendpunct}{\mcitedefaultseppunct}\relax
\EndOfBibitem
\bibitem[Nakao \latin{et~al.}(2010)Nakao, Kaczorowski, Wang, Cardinal,
  Buchholz, Sugimoto, Tobita, Lee, Toyoda, Billiar, \latin{et~al.}
  others]{nakao:2010}
Nakao,~A.; Kaczorowski,~D.~J.; Wang,~Y.; Cardinal,~J.~S.; Buchholz,~B.~M.;
  Sugimoto,~R.; Tobita,~K.; Lee,~S.; Toyoda,~Y.; Billiar,~T.~R. \latin{et~al.}
  Amelioration of rat cardiac cold ischemia/reperfusion injury with inhaled
  hydrogen or carbon monoxide, or both. \emph{J. Heart Lung Transp.}
  \textbf{2010}, \emph{29}, 544--553\relax
\mciteBstWouldAddEndPuncttrue
\mciteSetBstMidEndSepPunct{\mcitedefaultmidpunct}
{\mcitedefaultendpunct}{\mcitedefaultseppunct}\relax
\EndOfBibitem
\bibitem[Orcutt(1963)]{orcutt:1963}
Orcutt,~R.~H. Influence of Molecular Quadrupole Moments on the Second Virial
  Coefficient. \emph{J. Chem. Phys.} \textbf{1963}, \emph{39}, 605--608\relax
\mciteBstWouldAddEndPuncttrue
\mciteSetBstMidEndSepPunct{\mcitedefaultmidpunct}
{\mcitedefaultendpunct}{\mcitedefaultseppunct}\relax
\EndOfBibitem
\bibitem[Maroulis(2001)]{maroulis:2001}
Maroulis,~G. Accurate higher electric multipole moments for carbon monoxide.
  \emph{Chem. Phys. Lett.} \textbf{2001}, \emph{334}, 214--219\relax
\mciteBstWouldAddEndPuncttrue
\mciteSetBstMidEndSepPunct{\mcitedefaultmidpunct}
{\mcitedefaultendpunct}{\mcitedefaultseppunct}\relax
\EndOfBibitem
\bibitem[Brooks \latin{et~al.}(2009)Brooks, Brooks~III, MacKerell~Jr., Nilsson,
  Petrella, Roux, Won, Archontis, Bartels, Boresch, Caflisch, Caves, Cui,
  Dinner, Feig, Fischer, Gao, Hodoscek, Im, Kuczera, Lazaridis, Ma,
  Ovchinnikov, Paci, Pastor, Post, Schaefer, Tidor, Venable, Woodcock, Wu,
  Yang, York, and Karplus]{Charmm-Brooks-2009}
Brooks,~B.~R.; Brooks~III,~C.~L.; MacKerell~Jr.,~A.~D.; Nilsson,~L.;
  Petrella,~R.~J.; Roux,~B.; Won,~Y.; Archontis,~G.; Bartels,~C.; Boresch,~S.
  \latin{et~al.}  CHARMM: The Biomolecular Simulation Program. \emph{J. Comp.
  Chem.} \textbf{2009}, \emph{30}, 1545--1614\relax
\mciteBstWouldAddEndPuncttrue
\mciteSetBstMidEndSepPunct{\mcitedefaultmidpunct}
{\mcitedefaultendpunct}{\mcitedefaultseppunct}\relax
\EndOfBibitem
\bibitem[Hwang and {\it et al.}(2024)Hwang, and {\it et al.}]{MM.charmm:2024}
Hwang,~W.; {\it et al.}, CHARMM at 45: Enhancements in accessibility,
  functionality, and speed. \emph{J. Phys. Chem. B} \textbf{2024}, \emph{in
  print}, in print\relax
\mciteBstWouldAddEndPuncttrue
\mciteSetBstMidEndSepPunct{\mcitedefaultmidpunct}
{\mcitedefaultendpunct}{\mcitedefaultseppunct}\relax
\EndOfBibitem
\bibitem[Kuriyan \latin{et~al.}(1986)Kuriyan, Wilz, Karplus, and
  Petsko]{kuriyan:1986}
Kuriyan,~J.; Wilz,~S.; Karplus,~M.; Petsko,~G.~A. X-ray structure and
  refinement of carbon-monoxy (Fe II)-myoglobin at 1.5 {\AA} resolution.
  \emph{J. Mol. Biol.} \textbf{1986}, \emph{192}, 133--154\relax
\mciteBstWouldAddEndPuncttrue
\mciteSetBstMidEndSepPunct{\mcitedefaultmidpunct}
{\mcitedefaultendpunct}{\mcitedefaultseppunct}\relax
\EndOfBibitem
\bibitem[Plattner and Meuwly(2012)Plattner, and Meuwly]{MM.mb:2009}
Plattner,~N.; Meuwly,~M. Quantifying the importance of protein conformation on
  ligand migration in myoglobin. \emph{Biophys. J.} \textbf{2012}, \emph{102},
  333--341\relax
\mciteBstWouldAddEndPuncttrue
\mciteSetBstMidEndSepPunct{\mcitedefaultmidpunct}
{\mcitedefaultendpunct}{\mcitedefaultseppunct}\relax
\EndOfBibitem
\bibitem[Jorgensen \latin{et~al.}(1983)Jorgensen, Chandrasekhar, Madura, Impey,
  and Klein]{jorgensen-jcp-83}
Jorgensen,~W.~L.; Chandrasekhar,~J.; Madura,~J.~D.; Impey,~R.~W.; Klein,~M.~L.
  Comparison of simple potential functions for simulating liquid water.
  \emph{J. Chem. Phys.} \textbf{1983}, \emph{79}, 926--935\relax
\mciteBstWouldAddEndPuncttrue
\mciteSetBstMidEndSepPunct{\mcitedefaultmidpunct}
{\mcitedefaultendpunct}{\mcitedefaultseppunct}\relax
\EndOfBibitem
\bibitem[Gunsteren and Berendsen(1997)Gunsteren, and
  Berendsen]{SHAKE-Gunsteren-1997}
Gunsteren,~W.~V.; Berendsen,~H. Algorithms for Macromolecular Dynamics and
  Constraint Dynamics. \emph{Mol. Phys.} \textbf{1997}, \emph{34},
  1311--1327\relax
\mciteBstWouldAddEndPuncttrue
\mciteSetBstMidEndSepPunct{\mcitedefaultmidpunct}
{\mcitedefaultendpunct}{\mcitedefaultseppunct}\relax
\EndOfBibitem
\bibitem[Steinbach and Brooks(1994)Steinbach, and Brooks]{Steinbach1994}
Steinbach,~P.~J.; Brooks,~B.~R. New Spherical-Cutoff Methods for Long-Range
  Forces in Macromolecular Simulation. \emph{J. Comput. Chem.} \textbf{1994},
  \emph{15}, 667--683\relax
\mciteBstWouldAddEndPuncttrue
\mciteSetBstMidEndSepPunct{\mcitedefaultmidpunct}
{\mcitedefaultendpunct}{\mcitedefaultseppunct}\relax
\EndOfBibitem
\bibitem[Hoover(1985)]{Hoover1985}
Hoover,~W.~G. Canonical dynamics: Equilibrium phase-space distributions.
  \emph{Phys. Rev. A} \textbf{1985}, \emph{31}, 1695--1697\relax
\mciteBstWouldAddEndPuncttrue
\mciteSetBstMidEndSepPunct{\mcitedefaultmidpunct}
{\mcitedefaultendpunct}{\mcitedefaultseppunct}\relax
\EndOfBibitem
\bibitem[Ram\'{i}rez \latin{et~al.}(2004)Ram\'{i}rez, L\'{o}pez-Ciudad,
  Kumar~P, and Marx]{marx:2004}
Ram\'{i}rez,~R.; L\'{o}pez-Ciudad,~T.; Kumar~P,~P.; Marx,~D. Quantum
  corrections to classical time-correlation functions: Hydrogen bonding and
  anharmonic floppy modes. \emph{J. Chem. Phys.} \textbf{2004}, \emph{121},
  3973--3983\relax
\mciteBstWouldAddEndPuncttrue
\mciteSetBstMidEndSepPunct{\mcitedefaultmidpunct}
{\mcitedefaultendpunct}{\mcitedefaultseppunct}\relax
\EndOfBibitem
\bibitem[Guvench \latin{et~al.}(2011)Guvench, Mallajosyula, Raman, Hatcher,
  Vanommeslaeghe, Foster, Jamison, and MacKerell]{CHARMMFF-ALL36-Guvench2011}
Guvench,~O.; Mallajosyula,~S.~S.; Raman,~E.~P.; Hatcher,~E.;
  Vanommeslaeghe,~K.; Foster,~T.~J.; Jamison,~F.~W.,~II; MacKerell,~A.~D.,~Jr.
  CHARMM Additive All-Atom Force Field for Carbohydrate Derivatives and Its
  Utility in Polysaccharide and Carbohydrate-Protein Modeling. \emph{J. Chem.
  Theo. Comp.} \textbf{2011}, \emph{7}, 3162--3180\relax
\mciteBstWouldAddEndPuncttrue
\mciteSetBstMidEndSepPunct{\mcitedefaultmidpunct}
{\mcitedefaultendpunct}{\mcitedefaultseppunct}\relax
\EndOfBibitem
\bibitem[Jorgensen \latin{et~al.}(1983)Jorgensen, Chandrasekhar, Madura, Impey,
  and Klein]{TIP3P-Jorgensen-1983}
Jorgensen,~W.~L.; Chandrasekhar,~J.; Madura,~J.~D.; Impey,~R.~W.; Klein,~M.~L.
  Comparison of Simple Potential Functions for Simulating Liquid Water.
  \emph{J. Chem. Phys.} \textbf{1983}, \emph{79}, 926--935\relax
\mciteBstWouldAddEndPuncttrue
\mciteSetBstMidEndSepPunct{\mcitedefaultmidpunct}
{\mcitedefaultendpunct}{\mcitedefaultseppunct}\relax
\EndOfBibitem
\bibitem[Savelyev and MacKerell(2015)Savelyev, and MacKerell]{mackerell:2015}
Savelyev,~A.; MacKerell,~A. D.~J. Competition among Li$^+$, Na$^+$, K$^+$, and
  Rb$^+$ Monovalent Ions for DNA in Molecular Dynamics Simulations Using the
  Additive CHARMM36 and Drude Polarizable Force Fields. \emph{J. Phys. Chem. B}
  \textbf{2015}, \emph{119}, 4428--4440\relax
\mciteBstWouldAddEndPuncttrue
\mciteSetBstMidEndSepPunct{\mcitedefaultmidpunct}
{\mcitedefaultendpunct}{\mcitedefaultseppunct}\relax
\EndOfBibitem
\bibitem[Frisch \latin{et~al.}(2016)Frisch, Trucks, Schlegel, Scuseria, Robb,
  Cheeseman, Scalmani, Barone, Petersson, Nakatsuji, Li, Caricato, Marenich,
  Bloino, Janesko, Gomperts, Mennucci, Hratchian, Ortiz, Izmaylov, Sonnenberg,
  Williams-Young, Ding, Lipparini, Egidi, Goings, Peng, Petrone, Henderson,
  Ranasinghe, Zakrzewski, Gao, Rega, Zheng, Liang, Hada, Ehara, Toyota, Fukuda,
  Hasegawa, Ishida, Nakajima, Honda, Kitao, Nakai, Vreven, Throssell,
  Montgomery, Peralta, Ogliaro, Bearpark, Heyd, Brothers, Kudin, Staroverov,
  Keith, Kobayashi, Normand, Raghavachari, Rendell, Burant, Iyengar, Tomasi,
  Cossi, Millam, Klene, Adamo, Cammi, Ochterski, Martin, Morokuma, Farkas,
  Foresman, and Fox]{g09}
Frisch,~M.~J.; Trucks,~G.~W.; Schlegel,~H.~B.; Scuseria,~G.~E.; Robb,~M.~A.;
  Cheeseman,~J.~R.; Scalmani,~G.; Barone,~V.; Petersson,~G.~A.; Nakatsuji,~H.
  \latin{et~al.}  Gaussian˜16 {R}evision {C}.09. 2016; Gaussian Inc.
  Wallingford CT\relax
\mciteBstWouldAddEndPuncttrue
\mciteSetBstMidEndSepPunct{\mcitedefaultmidpunct}
{\mcitedefaultendpunct}{\mcitedefaultseppunct}\relax
\EndOfBibitem
\bibitem[Wang \latin{et~al.}(2021)Wang, Hou, and Heinz]{h2lj:2021}
Wang,~S.; Hou,~K.; Heinz,~H. Accurate and Compatible Force Fields for Molecular
  Oxygen, Nitrogen, and Hydrogen to Simulate Gases, Electrolytes, and
  Heterogeneous Interfaces. \emph{J. Chem. Theo. Comp.} \textbf{2021},
  \emph{17}, 5198--5213\relax
\mciteBstWouldAddEndPuncttrue
\mciteSetBstMidEndSepPunct{\mcitedefaultmidpunct}
{\mcitedefaultendpunct}{\mcitedefaultseppunct}\relax
\EndOfBibitem
\bibitem[Stoicheff(1957)]{stoicheff:1957}
Stoicheff,~B.~P. High Resolution Raman Spectroscopy of gases: IX. Spectra of
  H$_2$, HD, and D$_2$. \emph{Can. J. Phys.} \textbf{1957}, \emph{35},
  730--741\relax
\mciteBstWouldAddEndPuncttrue
\mciteSetBstMidEndSepPunct{\mcitedefaultmidpunct}
{\mcitedefaultendpunct}{\mcitedefaultseppunct}\relax
\EndOfBibitem
\bibitem[Neese \latin{et~al.}(2020)Neese, Wennmohs, Becker, and
  Riplinger]{orca:2020}
Neese,~F.; Wennmohs,~F.; Becker,~U.; Riplinger,~C. The ORCA quantum chemistry
  program package. \emph{J. Chem. Phys.} \textbf{2020}, \emph{152},
  224108\relax
\mciteBstWouldAddEndPuncttrue
\mciteSetBstMidEndSepPunct{\mcitedefaultmidpunct}
{\mcitedefaultendpunct}{\mcitedefaultseppunct}\relax
\EndOfBibitem
\bibitem[Weigend and Ahlrichs(2005)Weigend, and Ahlrichs]{ahlrichs:2005}
Weigend,~F.; Ahlrichs,~R. Balanced basis sets of split valence{,} triple zeta
  valence and quadruple zeta valence quality for H to Rn: Design and assessment
  of accuracy. \emph{Phys. Chem. Chem. Phys.} \textbf{2005}, \emph{7},
  3297--3305\relax
\mciteBstWouldAddEndPuncttrue
\mciteSetBstMidEndSepPunct{\mcitedefaultmidpunct}
{\mcitedefaultendpunct}{\mcitedefaultseppunct}\relax
\EndOfBibitem
\bibitem[Weigend(2006)]{weigend:2006}
Weigend,~F. Accurate Coulomb-fitting basis sets for H to Rn. \emph{Phys. Chem.
  Chem. Phys.} \textbf{2006}, \emph{8}, 1057--1065\relax
\mciteBstWouldAddEndPuncttrue
\mciteSetBstMidEndSepPunct{\mcitedefaultmidpunct}
{\mcitedefaultendpunct}{\mcitedefaultseppunct}\relax
\EndOfBibitem
\bibitem[Caldeweyher \latin{et~al.}(2019)Caldeweyher, Ehlert, Hansen,
  Neugebauer, Spicher, Bannwarth, and Grimme]{grimme:d4}
Caldeweyher,~E.; Ehlert,~S.; Hansen,~A.; Neugebauer,~H.; Spicher,~S.;
  Bannwarth,~C.; Grimme,~S. A generally applicable atomic-charge dependent
  London dispersion correction. \emph{J. Chem. Phys.} \textbf{2019},
  \emph{150}, 154122\relax
\mciteBstWouldAddEndPuncttrue
\mciteSetBstMidEndSepPunct{\mcitedefaultmidpunct}
{\mcitedefaultendpunct}{\mcitedefaultseppunct}\relax
\EndOfBibitem
\bibitem[Unke \latin{et~al.}(2017)Unke, Devereux, and Meuwly]{MM.mdcm:2017}
Unke,~O.~T.; Devereux,~M.; Meuwly,~M. Minimal distributed charges: Multipolar
  quality at the cost of point charge electrostatics. \emph{J. Chem. Phys.}
  \textbf{2017}, \emph{147}, 161712\relax
\mciteBstWouldAddEndPuncttrue
\mciteSetBstMidEndSepPunct{\mcitedefaultmidpunct}
{\mcitedefaultendpunct}{\mcitedefaultseppunct}\relax
\EndOfBibitem
\bibitem[Bondi(1964)]{BON:JPC64}
Bondi,~A. van der Waals Volumes and Radii. \emph{J. Phys. Chem.} \textbf{1964},
  \emph{68}, 441--451\relax
\mciteBstWouldAddEndPuncttrue
\mciteSetBstMidEndSepPunct{\mcitedefaultmidpunct}
{\mcitedefaultendpunct}{\mcitedefaultseppunct}\relax
\EndOfBibitem
\bibitem[Lumry \latin{et~al.}(1971)Lumry, Keyes, and Falley]{lumry:1971}
Lumry,~R.; Keyes,~M.~H.; Falley,~M. Heme proteins. II. Preparation and
  thermodynamic properties of sperm whale myoglobin. \emph{J. Am. Chem. Soc.}
  \textbf{1971}, \emph{93}, 2035--2040\relax
\mciteBstWouldAddEndPuncttrue
\mciteSetBstMidEndSepPunct{\mcitedefaultmidpunct}
{\mcitedefaultendpunct}{\mcitedefaultseppunct}\relax
\EndOfBibitem
\bibitem[Nutt \latin{et~al.}(2005)Nutt, Karplus, and Meuwly]{MM.mbno:2005}
Nutt,~D.~R.; Karplus,~M.; Meuwly,~M. Potential energy surface and molecular
  dynamics of MbNO: existence of an unsuspected FeON minimum. \emph{J. Phys.
  Chem. B} \textbf{2005}, \emph{109}, 21118--21125\relax
\mciteBstWouldAddEndPuncttrue
\mciteSetBstMidEndSepPunct{\mcitedefaultmidpunct}
{\mcitedefaultendpunct}{\mcitedefaultseppunct}\relax
\EndOfBibitem
\bibitem[Danielsson and Meuwly(2007)Danielsson, and Meuwly]{MM.mbcn:2007}
Danielsson,~J.; Meuwly,~M. Energetics and dynamics in MbCN: CN--vibrational
  relaxation from molecular dynamics simulations. \emph{J. Phys. Chem. B}
  \textbf{2007}, \emph{111}, 218--226\relax
\mciteBstWouldAddEndPuncttrue
\mciteSetBstMidEndSepPunct{\mcitedefaultmidpunct}
{\mcitedefaultendpunct}{\mcitedefaultseppunct}\relax
\EndOfBibitem
\bibitem[Lim \latin{et~al.}(1995)Lim, Jackson, and Anfinrud]{Lim:1995}
Lim,~M.; Jackson,~T.~A.; Anfinrud,~P.~A. Mid-infrared vibrational spectrum of
  CO after photodissociation from heme: evidence of a docking site in the heme
  pocket of hemoglobin and myoglobin. \emph{J. Chem. Phys.} \textbf{1995},
  \emph{102}, 4355\relax
\mciteBstWouldAddEndPuncttrue
\mciteSetBstMidEndSepPunct{\mcitedefaultmidpunct}
{\mcitedefaultendpunct}{\mcitedefaultseppunct}\relax
\EndOfBibitem
\bibitem[Nutt and Meuwly(2003)Nutt, and Meuwly]{MM.mbco:2003}
Nutt,~D.~R.; Meuwly,~M. Theoretical investigation of infrared spectra and
  pocket dynamics of photodissociated carbonmonoxy myoglobin. \emph{Biophys.
  J.} \textbf{2003}, \emph{85}, 3612--3623\relax
\mciteBstWouldAddEndPuncttrue
\mciteSetBstMidEndSepPunct{\mcitedefaultmidpunct}
{\mcitedefaultendpunct}{\mcitedefaultseppunct}\relax
\EndOfBibitem
\bibitem[Plattner and Meuwly(2008)Plattner, and Meuwly]{MM.mbco:2008}
Plattner,~N.; Meuwly,~M. The Role of Higher {CO}-Multipole Moments in
  Understanding the Dynamics of Photodissociated Carbonmonoxide in Myoglobin.
  \emph{Biophys. J.} \textbf{2008}, \emph{94}, 2505--2515\relax
\mciteBstWouldAddEndPuncttrue
\mciteSetBstMidEndSepPunct{\mcitedefaultmidpunct}
{\mcitedefaultendpunct}{\mcitedefaultseppunct}\relax
\EndOfBibitem
\bibitem[Bossa \latin{et~al.}(2005)Bossa, Amadei, Daidone, Anselmi, Vallone,
  Brunori, and {Di Nola}]{bossa:2005}
Bossa,~C.; Amadei,~A.; Daidone,~I.; Anselmi,~M.; Vallone,~B.; Brunori,~M.; {Di
  Nola},~A. Molecular Dynamics Simulation of Sperm Whale Myoglobin: Effects of
  Mutations and Trapped CO on the Structure and Dynamics of Cavities.
  \emph{Biophys. J.} \textbf{2005}, \emph{89}, 465--474\relax
\mciteBstWouldAddEndPuncttrue
\mciteSetBstMidEndSepPunct{\mcitedefaultmidpunct}
{\mcitedefaultendpunct}{\mcitedefaultseppunct}\relax
\EndOfBibitem
\bibitem[Banushkina and Meuwly(2007)Banushkina, and Meuwly]{MM.rough:2007}
Banushkina,~P.; Meuwly,~M. Diffusive dynamics on multidimensional rough free
  energy surfaces. \emph{J. Chem. Phys.} \textbf{2007}, \emph{127}, 13501\relax
\mciteBstWouldAddEndPuncttrue
\mciteSetBstMidEndSepPunct{\mcitedefaultmidpunct}
{\mcitedefaultendpunct}{\mcitedefaultseppunct}\relax
\EndOfBibitem
\bibitem[Banushkina and Meuwly(2005)Banushkina, and Meuwly]{MM.mbco:2005}
Banushkina,~P.; Meuwly,~M. Free-energy barriers in {MbCO} rebinding. \emph{J.
  Phys. Chem. B} \textbf{2005}, \emph{109}, 16911--16917\relax
\mciteBstWouldAddEndPuncttrue
\mciteSetBstMidEndSepPunct{\mcitedefaultmidpunct}
{\mcitedefaultendpunct}{\mcitedefaultseppunct}\relax
\EndOfBibitem
\bibitem[Lim \latin{et~al.}(1997)Lim, Jackson, and Anfinrud]{Lim:1997}
Lim,~M.; Jackson,~T.~A.; Anfinrud,~P.~A. Ultrafast rotation and trapping of
  carbon monoxide dissociated from myoglobin. \emph{Nature Structural Biology}
  \textbf{1997}, \emph{4}, 209--214\relax
\mciteBstWouldAddEndPuncttrue
\mciteSetBstMidEndSepPunct{\mcitedefaultmidpunct}
{\mcitedefaultendpunct}{\mcitedefaultseppunct}\relax
\EndOfBibitem
\bibitem[Merchant \latin{et~al.}(2003)Merchant, Noid, Thompson, Akiyama,
  Loring, and Fayer]{Merchant03}
Merchant,~K.~A.; Noid,~W.~G.; Thompson,~D.~E.; Akiyama,~R.; Loring,~R.~F.;
  Fayer,~M.~D. Structural assignments and dynamics of the A substates of MbCO:
  spectrally resolved vibrational echo experiments and molecular dynamics
  simulations. \emph{J. Phys. Chem. B} \textbf{2003}, \emph{107}, 4--7\relax
\mciteBstWouldAddEndPuncttrue
\mciteSetBstMidEndSepPunct{\mcitedefaultmidpunct}
{\mcitedefaultendpunct}{\mcitedefaultseppunct}\relax
\EndOfBibitem
\bibitem[Meuwly(2006)]{MM.mbco:2006}
Meuwly,~M. On the Influence of the Local Environment on the CO Stretching
  Frequencies in Native Myoglobin: Assignment of the B-States in MbCO.
  \emph{Chem. Phys. Chem.} \textbf{2006}, \emph{7}, 2061--2063\relax
\mciteBstWouldAddEndPuncttrue
\mciteSetBstMidEndSepPunct{\mcitedefaultmidpunct}
{\mcitedefaultendpunct}{\mcitedefaultseppunct}\relax
\EndOfBibitem
\bibitem[Raab(1998)]{raab:1998}
Raab,~C. G. D. I.~R. Measurement of the electric quadrupole moments of CO$_2$,
  CO, N2, Cl$_2$ and BF$_3$. \emph{Mol. Phys.} \textbf{1998}, \emph{93},
  49--56\relax
\mciteBstWouldAddEndPuncttrue
\mciteSetBstMidEndSepPunct{\mcitedefaultmidpunct}
{\mcitedefaultendpunct}{\mcitedefaultseppunct}\relax
\EndOfBibitem
\bibitem[Teal and MacWood(1935)Teal, and MacWood]{teal:1935}
Teal,~G.~K.; MacWood,~G.~E. The Raman spectra of the isotopic molecules H$_2$,
  HD, and D$_2$. \emph{J. Chem. Phys.} \textbf{1935}, \emph{3}, 760--764\relax
\mciteBstWouldAddEndPuncttrue
\mciteSetBstMidEndSepPunct{\mcitedefaultmidpunct}
{\mcitedefaultendpunct}{\mcitedefaultseppunct}\relax
\EndOfBibitem
\bibitem[Hu \latin{et~al.}(1996)Hu, Smith, and Spiro]{hu:1996}
Hu,~S.; Smith,~K.~M.; Spiro,~T.~G. Assignment of protoheme resonance Raman
  spectrum by heme labeling in myoglobin. \emph{J. Am. Chem. Soc.}
  \textbf{1996}, \emph{118}, 12638--12646\relax
\mciteBstWouldAddEndPuncttrue
\mciteSetBstMidEndSepPunct{\mcitedefaultmidpunct}
{\mcitedefaultendpunct}{\mcitedefaultseppunct}\relax
\EndOfBibitem
\end{mcitethebibliography}


\providecommand{\latin}[1]{#1}
\makeatletter
\providecommand{\doi}
  {\begingroup\let\do\@makeother\dospecials
  \catcode`\{=1 \catcode`\}=2 \doi@aux}
\providecommand{\doi@aux}[1]{\endgroup\texttt{#1}}
\makeatother
\providecommand*\mcitethebibliography{\thebibliography}
\csname @ifundefined\endcsname{endmcitethebibliography}
  {\let\endmcitethebibliography\endthebibliography}{}
\begin{mcitethebibliography}{0}
\providecommand*\natexlab[1]{#1}
\providecommand*\mciteSetBstSublistMode[1]{}
\providecommand*\mciteSetBstMaxWidthForm[2]{}
\providecommand*\mciteBstWouldAddEndPuncttrue
  {\def\EndOfBibitem{\unskip.}}
\providecommand*\mciteBstWouldAddEndPunctfalse
  {\let\EndOfBibitem\relax}
\providecommand*\mciteSetBstMidEndSepPunct[3]{}
\providecommand*\mciteSetBstSublistLabelBeginEnd[3]{}
\providecommand*\EndOfBibitem{}
\mciteSetBstSublistMode{f}
\mciteSetBstMaxWidthForm{subitem}{(\alph{mcitesubitemcount})}
\mciteSetBstSublistLabelBeginEnd
  {\mcitemaxwidthsubitemform\space}
  {\relax}
  {\relax}

\end{mcitethebibliography}

\end{document}


\begin{table}
\begin{center}
\caption{
{\bf Pocket definition.}
The names of each cavity or pocket within Mb is listed along with amino acid residue and residue number, which are used to define the cavity center.}
\label{sitab_pocket_table}
\begin{tabularx}{\textwidth}{c|X}
    Pocket Label & Residues and ID\\\hline
    Xe1 & LEU89, HSD93, LEU104, PHE138, ILE142, TYR146\\
    Xe2 & LEU72, ILE107, SER108, LEU135, PHE138, ARG139\\
    Xe3 & TRP7, LEU76, GLY80, ALA134, LEU137, PHE138\\
    Xe4 & GLY25, ILE28, LEU29, GLY65, VAL68, LEU72\\
    B-state & GLY25, ILE28, LEU29 , LEU32, THR39, LYS42, PHE43, LEU61, HSD64, GLY65, THR67, VAL68, LEU69, ALA71, LEU72, LEU89, HSD93, ILE99, TYR103, LEU104, ILE107\\
    Pocket 6 & TRP7, LEU9, VAL10, LEU11, HSD12, VAL13, TRP14, ALA15, LYS16, VAL17, GLU18, HSD24, LEU69, LEU72, GLY73, LEU76, LYS77, ILE111, ILE112, LEU115, PHE123, MET131, ALA134, LEU135\\
    Pocket 7 & VAL10, VAL13, ALA15, TRP14, LYS16, VAL17, GLU18, VAL21, GLY25, HSD24, ASP27, ILE28, VAL68, LEU69, THR70, LEU72, GLY73, ILE107, SER108, ALA110, ILE111, ILE112, HSD113, VAL114, LEU115, MET131, LEU135\\
    Pocket 8 & TRP7, GLN8, LEU9, VAL10, LEU11, HSD12, VAL13, TRP14, ALA15, VAL17, GLU18, LEU69, THR70, ALA71, LEU72, GLY73, ALA74, ILE75, LEU76, LYS77, LYS78, LYS79, MET131, ALA134\\ 
    Pocket 9 & LEU9, VAL10, LEU11, HSD12, VAL13, TRP14, ALA15, LYS16, VAL17, GLU18, LEU69, LEU72, GLY73, LEU76, ILE111, ILE112, VAL114, LEU115, HSD116, HSD119, ASP122, PHE123, ALA127, GLN128, ALA130, MET131, ASN132, ALA134, LEU135\\    
\end{tabularx}
\end{center}
\end{table}

\begin{center}
\begin{figure}
    \centering
    \includegraphics[width=0.5\linewidth]{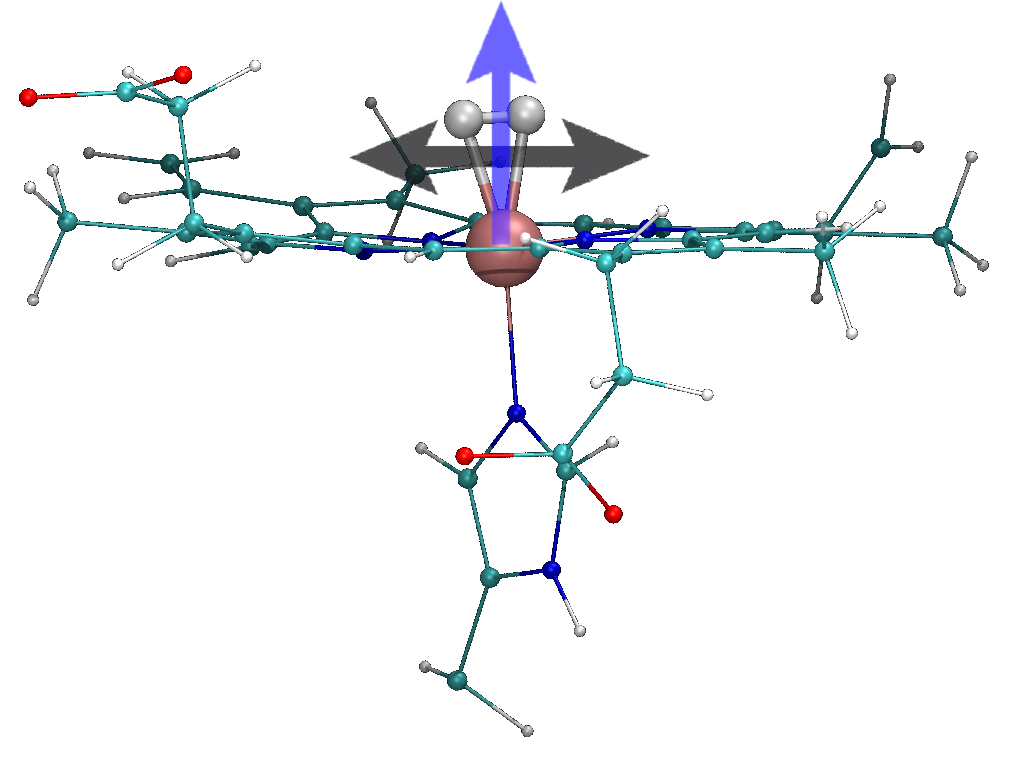}
    \caption{
    {\bf PES scan directions.}
    Illustration of the scan-directions to map out the potential energy
    surface of 
    H$_2$ interacting with the heme-unit of Mb (see Figure 2).
    The arrows indicate the lateral movement ($r-$direction in black) and the $z-$direction in blue.}
    \label{sifig_pes}
\end{figure}
\end{center}